\documentclass[pdftex,twocolumn,epjc3]{svjour3}          
\pdfoutput=1
\RequirePackage[T1]{fontenc}

\RequirePackage{graphicx}
\RequirePackage{mathptmx}      
\RequirePackage{flushend}
\RequirePackage[numbers,sort&compress]{natbib}
\RequirePackage[colorlinks,citecolor=blue,urlcolor=blue,linkcolor=blue]{hyperref}
\RequirePackage{amsmath}
\RequirePackage[british]{babel} 
\RequirePackage{bm}              
\RequirePackage{lineno}    
\RequirePackage[latin9]{inputenc} 
\RequirePackage{seqsplit}
\RequirePackage{xcolor}
\RequirePackage{textcomp}
\RequirePackage{amssymb}
\RequirePackage{booktabs}
\journalname{Eur. Phys. J. C}

\usepackage{widetext}
\newcommand{\DIC}{\Delta_{\rm IC}}

\begin{document}
\sloppy

\title{Impact of low-$x$ resummation on QCD analysis of HERA data}


\author{xFitter Developers' team:
     Hamed Abdolmaleki       \thanksref{a}
\and Valerio Bertone         \thanksref{m,b}
\and Daniel Britzger         \thanksref{dd}
\and Stefano Camarda         \thanksref{d}
\and Amanda Cooper-Sarkar    \thanksref{e}
\and Francesco Giuli         \thanksref{e}
\and Alexander Glazov        \thanksref{c}
\and Aleksander Kusina       \thanksref{g}
\and Agnieszka Luszczak      \thanksref{c,aa}
\and Fred Olness             \thanksref{h}
\and Andrey Sapronov         \thanksref{i}
\and Pavel Shvydkin          \thanksref{i}
\and Katarzyna Wichmann      \thanksref{c}
\and Oleksandr Zenaiev       \thanksref{c}
 and Marco Bonvini           \thanksref{l}}

\institute{Faculty of Physics, Semnan University, 35131-19111 Semnan,
  Iran \label{a}
  \and Department of Physics and Astronomy, VU University, NL-1081 HV
  Amsterdam, The Netherlands \label{m}
  \and NIKHEF Theory Group, Science Park 105, 1098 XG Amsterdam, The
  Netherlands \label{b}
  \and Physikalisches Institut, Universit{\" a}t Heidelberg, Im Neuenheimer Feld 226, 69120 Heidelberg, Germany \label{dd} 
  \and CERN, CH-1211 Geneva 23, Switzerland \label{d}
  \and Particle Physics, Denys Wilkinson Bdg, Keble Road,
  University of Oxford, OX1 3RH Oxford, UK \label{e}
  \and Deutsches Elektronen-Synchrotron (DESY), Notkestrasse 85,
  D-22607 Hamburg, Germany \label{c}
  \and Institute of Nuclear Physics, Polish Academy of Sciences,
  ul. Radzikowskiego 152, 31-342 Cracow, Poland \label{g}
  \and T. Kosciuszko Cracow University of Technology, PL-30-067, Cracow, Poland \label{aa}
  \and SMU Physics, Box 0175 Dallas, TX 75275-0175, United States of
  America \label{h}
  \and Joint Institute for Nuclear Research (JINR), Joliot-Curie 6,
  141980, Dubna, Moscow Region, Russia \label{i}
  \and INFN, sezione di Roma 1, Piazzale Aldo Moro 5, 00185 Rome, Italy \label{l}
}

\date{Received: date / Accepted: date}

\maketitle

\begin{abstract}
  Fits to the final combined HERA deep-inelastic scattering
  cross-section data within the conventional DGLAP framework of QCD
  have shown some tension at low $x$ and low $Q^2$.  A resolution of
  this tension incorporating $\ln(1/x)$-resummation terms into the
  HERAPDF fits is investigated using the {\tt xFitter} program. The
  kinematic region where this resummation is important is
  delineated. Such high-energy resummation not only gives a better
  description of the data, particularly of the longitudinal structure
  function $F_L$, it also results in a gluon PDF which is steeply
  rising at low $x$ for low scales, $Q^2 \simeq 2.5$~GeV$^2$, contrary
  to the fixed-order NLO and NNLO gluon PDF.
  \footnotetext{Preprint numbers: DESY 18-017, NIKHEF/2018-005\\
    Correspondence: {\tt valerio.bertone@cern.ch}}
\end{abstract}



\section{Introduction}

Deep-inelastic-scattering (DIS) experiments have traditionally been
used to probe the parton distribution functions (PDFs) of the
proton. A very broad range of resolving power, as characterised by
$Q^2$ (the negative four-momentum transfer squared) and by Bjorken $x$
(which is interpreted as the fraction of the proton's momentum taken
by the struck parton), was accessed at HERA. Perturbative quantum
chromo-dynamics (pQCD) is expected to describe these data, such that
PDFs can be extracted for $Q^2 \gtrsim 2$-$3$~GeV$^2$.

The final combined inclusive cross-section data from the HERA
experiments H1 and ZEUS~\cite{Abramowicz:2015mha} were input to QCD
analyses using fixed-order pQCD at LO, NLO and NNLO to provide the
HERAPDF2.0 set of parton distributions. However, some tension was
observed at low $Q^2$ such that the $\chi^2$ for these fits drops
steadily as the minimum energy $Q_{\rm min}^2$ of the data entering
the fit is raised up to $Q^2_{\rm min}\simeq 10$~GeV$^2$ (see Fig.~19
of Ref~\cite{Abramowicz:2015mha}). This turns out to be true for all
perturbative orders and is not mitigated by going to higher order. In
particular, the $\chi^2$ of the NNLO fits is not better than the NLO
fit for low values of $Q_{\rm min}^2$.

A further observation is that the increased $\chi^2$ of the fits to
the low-$Q^2$ data is largely attributable to the kinematic region of
low $x$ and high $y$ (where $y$ = $Q^2/sx$ and $\sqrt{s}$ is the
centre-of-mass energy) in the neutral-current reduced cross-section
data $\sigma_{\rm red}$, defined as\footnote{At low $Q^2$ and low $x$
  the parity-violating term proportional to $xF_3$ can be neglected.}:
\begin{equation}\label{eq:redxsec}
  \sigma_{\rm red} =F_2 - \frac{y^2}{Y_{+}} F_L\,,
\end{equation}
where, $F_2$ and $F_L$ are the structure functions, which are related
to the parton distributions~\cite{Gao:2017yyd}, and
$Y_{+}= 1 + (1-y)^2$. In this kinematic region (low $Q^2$ and low $x$)
the data take a turn-over (see Fig.~\ref{fig:data920} and
\textit{e.g.}\ Fig.~59 of Ref.~\cite{Abramowicz:2015mha}).  This
effect can be ascribed to the negative term proportional to $F_L$ in
Eq.~(\ref{eq:redxsec}). However, fits to data using fixed-order pQCD
do not describe this turn-over very well, suggesting that a larger
$F_L$ is needed for a better description. This in turn suggests that
the gluon evolution may need modification since $F_L$ is closely
related to it~\cite{CooperSarkar:1987ds}.

It has been noted that the addition of a higher-twist term to the
$F_L$ structure function improves the quality of the
fits~\cite{Abt:2016vjh,Harland-Lang:2016yfn,Alekhin:2016uxn,Motyka:2017xgk}. Such
a higher-twist term improves the $\chi^2$ both at NLO and NNLO, so
that the NNLO $\chi^2$ becomes better than the NLO one. Moreover, it
also improves the description both of the low-$x$, high-$y$ reduced
cross sections and the $F_L$ data from HERA~\cite{Andreev:2013vha}
(see Figs.~4, 5, and 11 of Ref.~\cite{Abt:2016vjh}).

Recently, an alternative approach which can improve the description of
low-$Q^2$ data has been proposed. Since the kinematics of HERA is such
that low-$Q^2$ data is also at low $x$, it has been suggested that the
DGLAP resummation of $\ln Q^2$ terms should be augmented by $\ln(1/x)$
(BFKL) resummation~\cite{Ball:2017otu}.  This idea is not new: the
necessary calculations have been explored in
Refs.~\cite{Ball:1995vc,Ball:1997vf,Altarelli:1999vw,Altarelli:2000mh,
  Altarelli:2001ji,Altarelli:2003hk,Altarelli:2005ni,Ball:2007ra,
  Altarelli:2008xp,Altarelli:2008aj},~\cite{Salam:1998tj,Ciafaloni:1999yw,
  Ciafaloni:2003rd,Ciafaloni:2007gf} and
\cite{Thorne:1999sg,Thorne:1999rb, Thorne:2001nr,White:2006yh}. They
also inspired various phenomenological fits, \textit{e.g.}
Ref.~\cite{Luszczak:2016bxd}.  However, a complete implementation,
such that these terms can readily be used for fitting PDFs, is new.
This was possible thanks to new theoretical developments and the
publication of the \texttt{HELL} code which implements $\ln(1/x)$
resummation~\cite{Bonvini:2016wki,Bonvini:2017ogt}.  The present paper
explores the implementation of the public {\tt HELL 2.0} code into
{\tt xFitter}~\cite{Alekhin:2014irh,h1zeus:2009wt,Aaron:2009kv} and
the consequences for a HERAPDF-style fit using this code.  The
conclusions of the study of Ref.~\cite{Ball:2017otu} are confirmed by
our analysis.  Having interfaced \texttt{HELL} to the public
\texttt{xFitter} tool makes $\ln(1/x)$ resummation accessible for any
future PDF determination.

\section{Input data sets}

The final combined $e^{\pm}p$ cross-section measurements of H1 and
ZEUS~\cite{Abramowicz:2015mha} cover the kinematic range of $Q^2$ from
$0.045$~GeV$^2$ to $50000$~GeV$^2$ and of Bjorken $x$ from $0.65$ down
to $6\cdot 10^{-7}$. There are 169 correlated sources of uncertainty
and total uncertainties are below 1.5$\%$ over the $Q^2$ range
$3$~GeV$^2 < Q^2 <500$~GeV$^2$ and below 3$\%$ up to
$Q^2=3000$~GeV$^2$. There are data from neutral-current (NC) and
charged-current (CC) processes and for $e^+p$ and $e^-p$
scattering. In addition to this, the NC $e^+p$ data are available for
several different proton beam energies. The availability of NC and CC
precision data over a large phase space allows for the determination
of PDFs. The difference between the NC $e^+p$ and $e^-p$ cross
sections at high $Q^2$, together with the high-$Q^2$ CC data,
constrains the valence distributions. The lower-$Q^2$ NC data
constrain the low-$x$ sea quark distributions and their precisely
measured $Q^2$ variation constrains the gluon
distribution. Furthermore, the inclusion of NC data at different beam
energies probes the longitudinal structure function $F_L$ such that
the gluon is further constrained.  A minimum $Q^2$ cut of
$Q^2 \ge 3.5$~GeV$^2$ is imposed on inclusive HERA data. This gives
1145 data points included in the fit.

In addition, HERA combined charm~\cite{Abramowicz:1900rp} and beauty
data~\cite{Aaron:2009af,Abramowicz:2014zub} from ZEUS and H1 are also
available.  Reduced cross sections for charm production cover the
kinematic range $2.5$~GeV$^2 < Q^2 < 2000$~GeV$^2$,
$3\cdot 10^{-5} < x < 0.05$. There are 48 correlated sources of
uncertainty and the total uncertainties are typically 6$\%$ at small
$x$ and medium $Q^2$ and 10$\%$ on average. There are 47 charm data
points included in the fit after the $Q^2$ cut of
$Q^2 \ge 3.5$~GeV$^2$ is imposed.  ZEUS reduced cross sections for
beauty cover the kinematic range $6.5$~GeV$^2 < Q^2 < 600$~GeV$^2$,
$1.5\cdot 10^{-4} < x < 0.035$.  There are 13 correlated sources of
uncertainty and the total uncertainties range from about 10$\%$ to
20$\%$. There are 17 ZEUS beauty data points included in the fit. H1
reduced cross sections for beauty cover the kinematic range
$5$~GeV$^2 < Q^2 < 2000$~GeV$^2$, $2\cdot 10^{-4} < x < 0.055$. There
are 14 correlated sources of uncertainty and the total uncertainties
range from about 20$\%$ to 40$\%$. There are 12 H1 beauty data points
included in the fit.  The inclusion of charm and beauty data in the
fit is useful to determine the optimal charm and beauty pole
masses. Additionally, since the charm data in particular extend to
rather small values of $x$, they may be sensitive to $\ln(1/x)$
resummation effects.

\section{Fit strategy}
\label{sec:setup}

The present QCD analysis uses the {\tt xFitter}
program~\cite{Alekhin:2014irh,h1zeus:2009wt,Aaron:2009kv} and is based
on the HERAPDF2.0 setup.  However, in order to facilitate the
inclusion of small-$x$ resummation, some differences have been
introduced with respect to the HERAPDF2.0 theory settings.  In this
section, we first present our setup, and then highlight the features
that differ from those of HERAPDF2.0.

In the present analysis, we use the {\tt APFEL}
code~\cite{Bertone:2013vaa} to compute the structure functions and the
solution of the DGLAP evolution equations. The {\tt APFEL} code
implements the FONLL variable-flavour-number
scheme~\cite{Forte:2010ta} for the treatment of heavy quarks.  The
heavy-quark pole masses were initially chosen to be $m_c=1.43$~GeV and
$m_b=4.5$~GeV.  These choices follow those of the HERAPDF2.0, but the
sensitivity to these values is reviewed in Sec.~\ref{sec:results}.
The choice of \texttt{APFEL} is motivated by the fact that it has been
interfaced to the \texttt{HELL} code which is needed to include
small-$x$ resummation.  The {\tt HELL} code is a standalone code that
implements the resummation corrections to the DGLAP splitting
functions $P$ and to the DIS coefficient functions $C$ (both massless
and massive) up to next-to-leading-log accuracy in $\ln(1/x)$
(NLL$x$).\footnote{The resummation procedure of the logarithmically
  enhanced terms of the massive coefficient functions implies the
  computation of the corrections due to $\ln(1/x)$ resummation on the
  PDF matching conditions. Details can be found in
  Ref~\cite{Bonvini:2017ogt}. These effects are also included in the
  results presented below.}  The output of \texttt{HELL} is in the
form of corrections $\Delta_l P^{{\rm N}^k{\rm LL}x}$ and
$\Delta_l C^{{\rm N}^k{\rm LL}x}$ to the fixed-order quantities
$P^{{\rm N}^l{\rm LO}}$ and $C^{{\rm N}^l{\rm LO}}$ (with $k=0,1$ and
$l=0,1,2$), which have to be supplied externally.  For example, the
expressions needed to compute the DGLAP evolution and the DIS
structure functions at NNLO+NLL$x$ accuracy are given by:
\begin{equation}
\begin{array}{l}
P = P^{{\rm N}^2{\rm LO}} + \Delta_2 P^{{\rm NLL}x}\,,\\
C = C^{{\rm N}^2{\rm LO}} + \Delta_2 C^{{\rm NLL}x}\,.
\end{array}
\end{equation}
In our case, the fixed-order contributions are those implemented in
{\tt APFEL}, which is then used in conjunction with {\tt HELL} to
compute DGLAP evolution and structure functions with the inclusion of
$\ln(1/x)$ resummation.

QCD evolution yields the PDFs at any value of $Q^2$ if they are
parameterised as functions of $x$ at some initial scale $Q^2_0$. This
scale is chosen to be $Q^2_0= 2.56$~GeV$^2$ (\textit{i.e.}
$Q_0=1.6$~GeV). The reason is that the numerical computation of
$\ln(1/x)$-resummation corrections may become unreliable at low scales
due to the large value of the strong coupling $\alpha_S$. As a
consequence, it is safer to keep the initial scale as high as
possible: the value $Q_0 = 1.6$~GeV represents a good compromise which
is known to lead to reliable results~\cite{Bonvini:2017ogt}.  Also, as
all data we are interested in lie above this scale, this choice does
not force us to exclude any of the interesting datapoints. However,
this choice gives us an initial scale that is above the default
charm-quark matching scale $\mu_c$, \textit{i.e.}\ the scale at which
the number of active flavours switches from $n_f=3$ to $n_f=4$,
usually taken to be equal to the charm pole mass
$m_c=1.47$~GeV$^2<Q_0$.  This could appear to be a problem since we
wish to fit just the light-quark PDFs and generate the heavy-quark
PDFs, including the charm PDF, dynamically.  However, since the
charm-quark matching scale $\mu_c$ is an unphysical scale, its value
can be mo\-di\-fied at will, provided it is kept close to $m_c$ to
avoid ge\-ne\-ra\-ting large logarithms.  Thus, we have used a feature
of the {\tt APFEL} code, discussed in Ref.~\cite{Bertone:2017ehk},
that allowed us to displace the charm-quark matching scale $\mu_c$
above $Q_0$ while keeping the charm mass fixed at $m_c = 1.43$~GeV. In
particular, we have chosen $\mu_c = 1.12\, m_c\simeq1.6$~GeV, which is
slightly larger than $Q_0$.
 
The quark distributions at the initial scale $Q_0^2$ are represented
by the generic form:
\begin{equation}
  xq_i(x,Q_0) = A_i x^{B_i} (1-x)^{C_i} P_i(x),
\label{eqn:pdf}
\end{equation}
where $P_i(x)$ defines a polynomial in positive powers of $x$. The
parametrised quark distributions $q_i$ are chosen to be the valence
quark distributions ($xu_v$, $xd_v$) and the light anti-quark
distributions ($x\bar{U}=x\bar{u}$, $x\bar{D}=x\bar{d}+x\bar{s}$). The
gluon distribution is parametrised with the more flexible form:
\begin{equation}\label{eqn:gluonpdf}
xg(x) = A_g x^{B_g} (1-x)^{C_g}P_g(x) - A'_g x^{B'_g} (1-x)^{C'_g}\,.
\end{equation}
The normalisation parameters $A_{u_v}$ and $A_{d_v}$ are fixed using
the quark counting rules and $A_g$ using the momentum sum rule. The
normalisation and slope parameters, $A$ and $B$, of $\bar{u}$ and
$\bar{d}$ are set equal such that $x\bar{u} = x\bar{d}$ at very small
$x$. The strange PDFs $xs$ and $x\bar{s}$ are parametrised as
$xs = x\bar{s}=0.4x\bar{D}$, representing a suppression of strangness
with respect to the light down-type sea quarks, but the input data are
not sensitive to the fraction of strangeness. Terms with positive
powers of $x$ are included in the polynomial $P_i(x)$ only if required
by the data, following the procedure described in
Ref.~\cite{h1zeus:2009wt}. This leads to the additional terms
$P_{u_v}(x)=1+ E_{u_v} x^2$ and $P_{\bar{U}}= 1 + D_{\bar{U}} x$ and
gives a total of 14 free parameters.\footnote{The exact value of
  $C_g'$ does not matter provided it is large enough that the negative
  term in Eq.~(\ref{eqn:gluonpdf}) does not contribute at large
  $x$. We take $C_g' = 25$ following the HERAPDF
  strategy~\cite{Abramowicz:2015mha}.}  The reference value of the
strong coupling constant is set to $\alpha_S(M_Z) = 0.118$.

The setting presented so far differs from the one of the HERAPDF2.0
analysis in some respects that we now highlight.
\begin{itemize}
\item For the HERAPDF2.0 analysis the DGLAP evolution and the
  light-quark coefficient functions are taken from the {\tt QCDNUM}
  code~\cite{Botje:2010ay} up to NNLO. There is no difference between
  the results of {\tt QCDNUM} and {\tt APFEL} for the treatment of
  light quarks. However, {\tt APFEL} implements the FONLL
  variable-flavour-number scheme~\cite{Forte:2010ta}, not the TR
  ``optimal'' variable-flavour-number scheme of
  Refs.~\cite{Thorne:1997ga,Thorne:2006qt}, which is the default in
  HERAPDF analyses.  The choice of the variable-flavour-number scheme
  represents the first main difference of the present analysis with
  respect to HERAPDF2.0.
\item A second difference is the scale at which PDFs are
  parameterised. In this analysis we have chosen $Q^2_0= 2.56$~GeV$^2$
  as compared to 1.9~GeV$^2$ of HERAPDF2.0.
\item Furthermore we have chosen the charm-quark matching scale as
  $\mu_c = 1.12 m_c\simeq1.6$~GeV, as explained above. This represents
  the third main difference with respect to the HERAPDF2.0 analysis.
  Note that, once this new setting is adopted, the optimal values of
  the charm and beauty masses may change and need to be reassessed
  (see Sect.~\ref{sec:results}), thus representing an extra (minor)
  difference with respect to HERAPDF2.0.
\end{itemize}
The impact of these differences will be investigated in
Sect.~\ref{sec:results} before including resummation effects.

\begin{table*}[t]
\begin{center}
\begin{tabular}{cccccc}
  \hline
  & Step-1& Step-2 &Step-3 & Step-4 & Step-5 \\
  \hline
  & HERAPDF2.0 &  TR$\to$FONLL-C  & raise the charm & raise the& include NLL$x$  \\
  &  NNLO      &                 & matching scale $\mu_c$ & initial scale $Q_0$ & resummation\\
  \hline
  HERA $\chi^2/\rm{d.o.f.}$ & 1363/1131  & 1387/1131 & 1390/1131 & 1388/1131& 1316/1131\\
  \hline
\end{tabular}
\caption{The $\chi^2$ per degree of freedom (d.o.f.) for PDF fits
  under different conditions, starting from the settings of the
  HERAPDF2.0 analysis at NNLO.}
\label{tab:fitresults1}
\end{center}
\end{table*}

\begin{figure*}[t]
  \begin{centering}
    \includegraphics[width=0.32\textwidth]{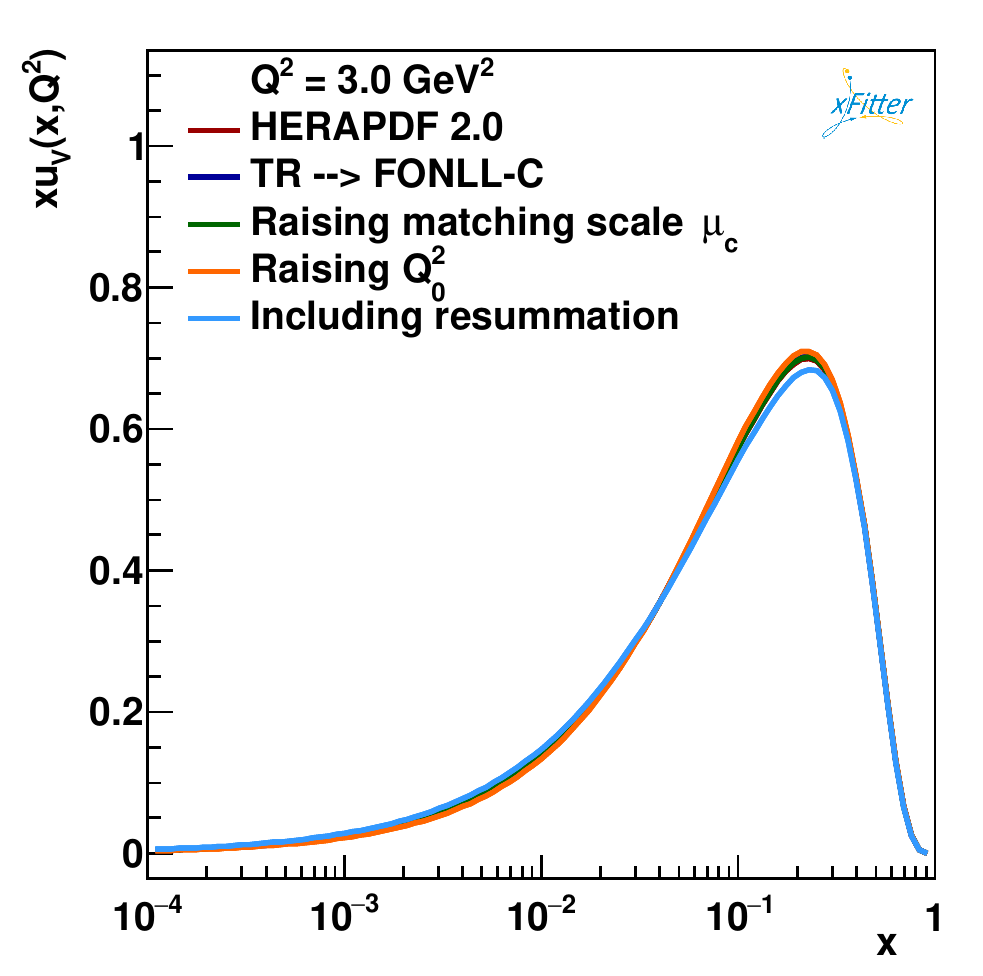}
    \includegraphics[width=0.32\textwidth]{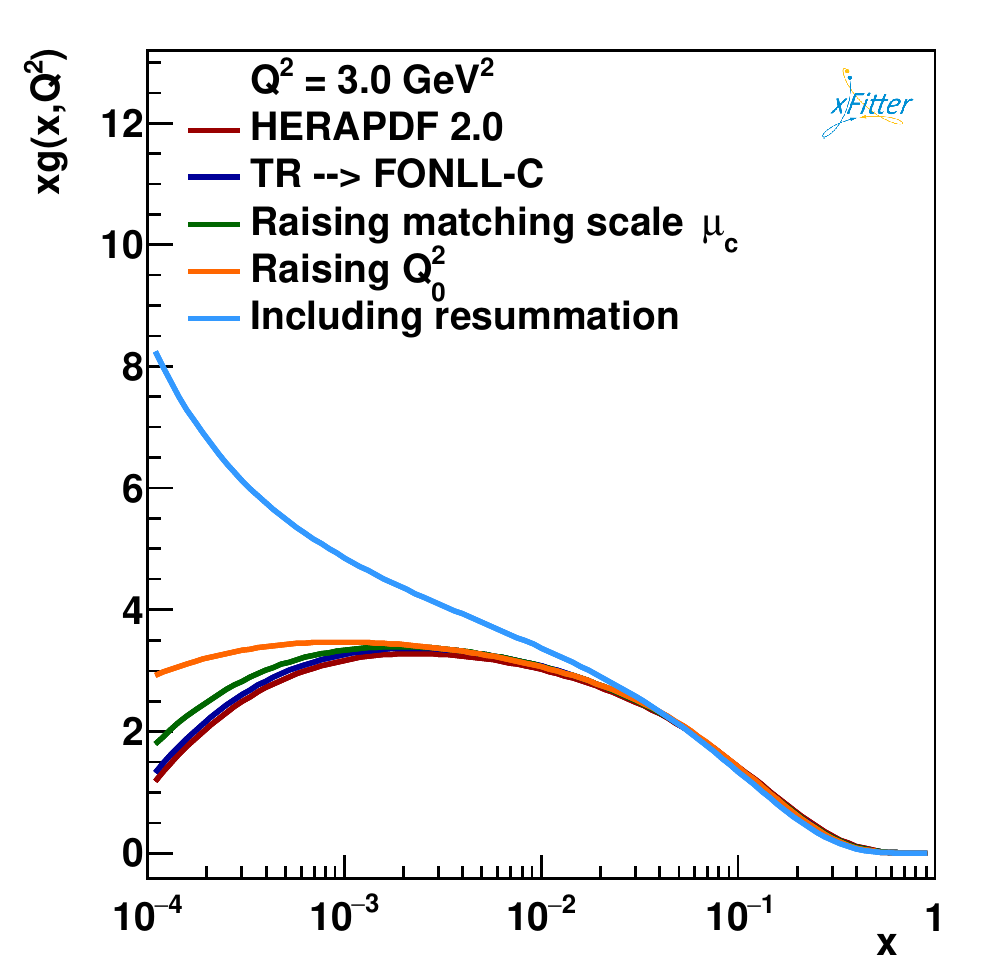}
    \includegraphics[width=0.32\textwidth]{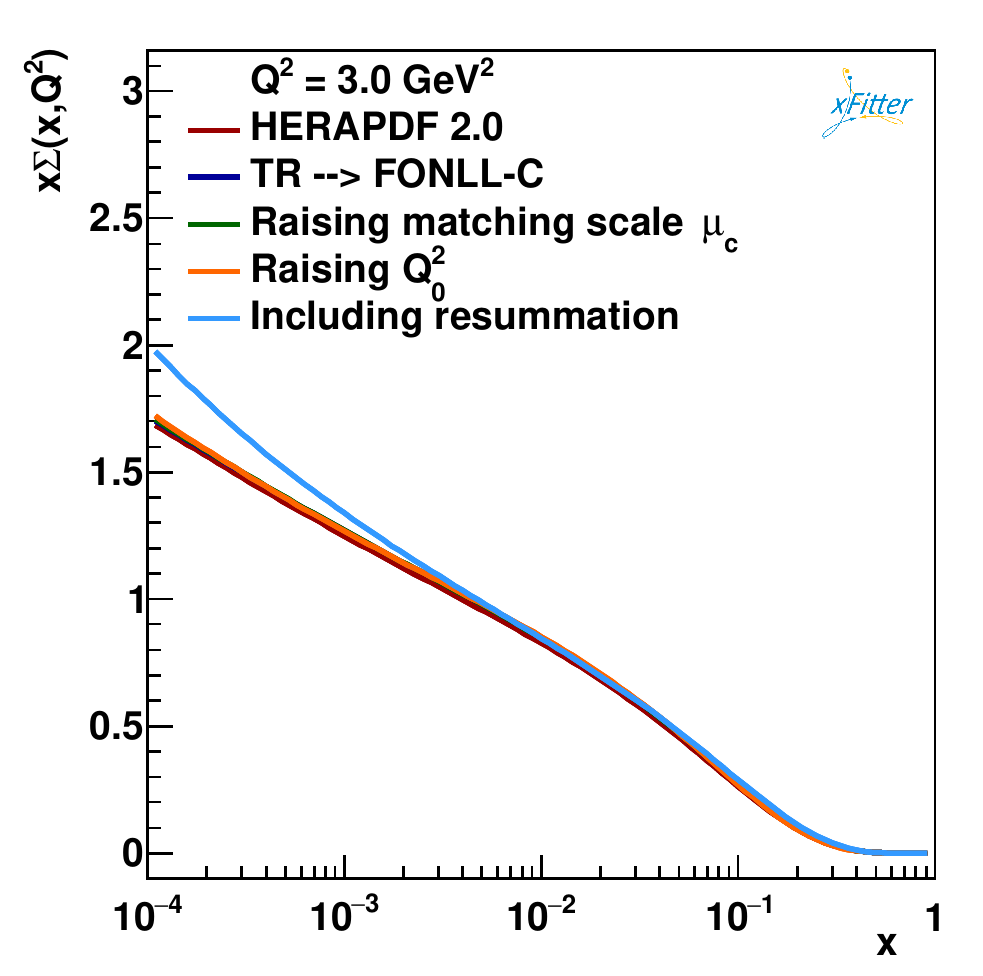}
    \caption{The up valence PDF $xu_v$, the gluon PDF $xg$ and the
      total singlet PDF $x\Sigma$ for each of the 5 steps outlined in
      the text.}
    \label{fig:5steps}
  \end{centering}
\end{figure*}

It is also useful to compare the present settings with those used in
the NNPDF3.1sx analysis~\cite{Ball:2017otu}.  Indeed, the same tools
have been used to compute structure functions and PDF evolution,
namely {\tt APFEL} and {\tt HELL}.  The differences, however, are
significant.  First, the fit methodology is very different, as NNPDF
uses a Monte Carlo approach with neural network parametrisation of
PDFs.  Second, the data considered in NNPDF3.1sx include several
additional DIS data sets from other experiments. In addition, that
analysis also includes Tevatron and LHC data.  Most importantly from
the theory point of view, there is a difference in the way charm is
treated.  In particular, in the NNPDF analysis the charm PDFs are
fitted to data. In this analysis, a more conventional approach is used
in which the charm PDFs are generated perturbatively. This approach in
the framework of the FONLL scheme allows for the inclusion of a
damping factor that suppresses subleading higher-order corrections
that might be significant at scales comparable to the charm
mass~\cite{Forte:2010ta}. We include this damping factor in our
computation as it turns out to improve dramatically the description of
the data at NNLO.  We will further comment on the effect of this
damping factor in Sect.~\ref{sec:nnpdf}.

\section{Results}
\label{sec:results}

The effect of $\ln(1/x)$ resummation on splitting functions and DIS
coefficient functions is more dramatic at NNLO than at
NLO~\cite{Bonvini:2017ogt}.  In fact, the full calculation with
NNLO+NLL$x$ resummation is closer to the NLO result than it is to the
NNLO result. This is not accidental and is mostly due to the
perturbative instability of the NNLO correction to the splitting
functions generated by small-$x$ logarithms~\cite{Ball:2017otu}. Thus,
to better assess the impact of the $\ln(1/x)$ resummation on the
original HERAPDF analysis, we only focus on NNLO fits.

\subsection{\textit{Transition to the new fit settings}}

Since the setup described in Sect.~\ref{sec:setup} differs in various
respects from that of the HERAPDF2.0 analysis, we first investigate
the effect of these changes in the determination of PDFs.  A
step-by-step approach is followed.  The changes in the fit quality are
summarised in Tab.~\ref{tab:fitresults1}, while the effect on the
$u_v$, the total singlet and the gluon PDFs\footnote {The $d_v$ is not
  shown since its shape is unchanged, just as for the $u_v$.}  is
shown in Fig.~\ref{fig:5steps}.

The starting point is the HERAPDF2.0 analysis, that has a $\chi^2$ of
1363 units for 1131 degrees of freedom (see Tab.~4 of
Ref.~\cite{Abramowicz:2015mha}) using only the HERA inclusive data.
First we move to the use the FONLL-C scheme~\cite{Forte:2010ta} in
place of the TR scheme~\cite{Thorne:1997ga,Thorne:2006qt}, with all
the same settings.  The PDFs look remarkably similar to the HERAPDF2.0
results as illustrated in Fig.~\ref{fig:5steps}.  However, the
$\chi^2$ increases significantly to 1387.  This was expected because
FONLL-C is known to lead to a worse description of the data than the
TR scheme at this order, as discussed in
Ref.~\cite{Abramowicz:2015mha} (see Fig.~20 of that reference).  The
origin of this deterioration is related to the details of the
construction of the observables within each scheme, which differs in
various respects, from the perturbative orders at which each
individual contribution is retained to the presence of
phenomenological smoothing functions.  A full assessment of these
differences and their importance for the description of HERA data is
beyond the scope of this paper.  Some considerations on the impact of
the details of the heavy-flavour scheme in our fits with and without
$\ln(1/x)$ resummation will be given in Sect.~\ref{sec:nnpdf}.

\begin{figure}[t]
  \begin{centering}
    \includegraphics[width=0.49\textwidth]{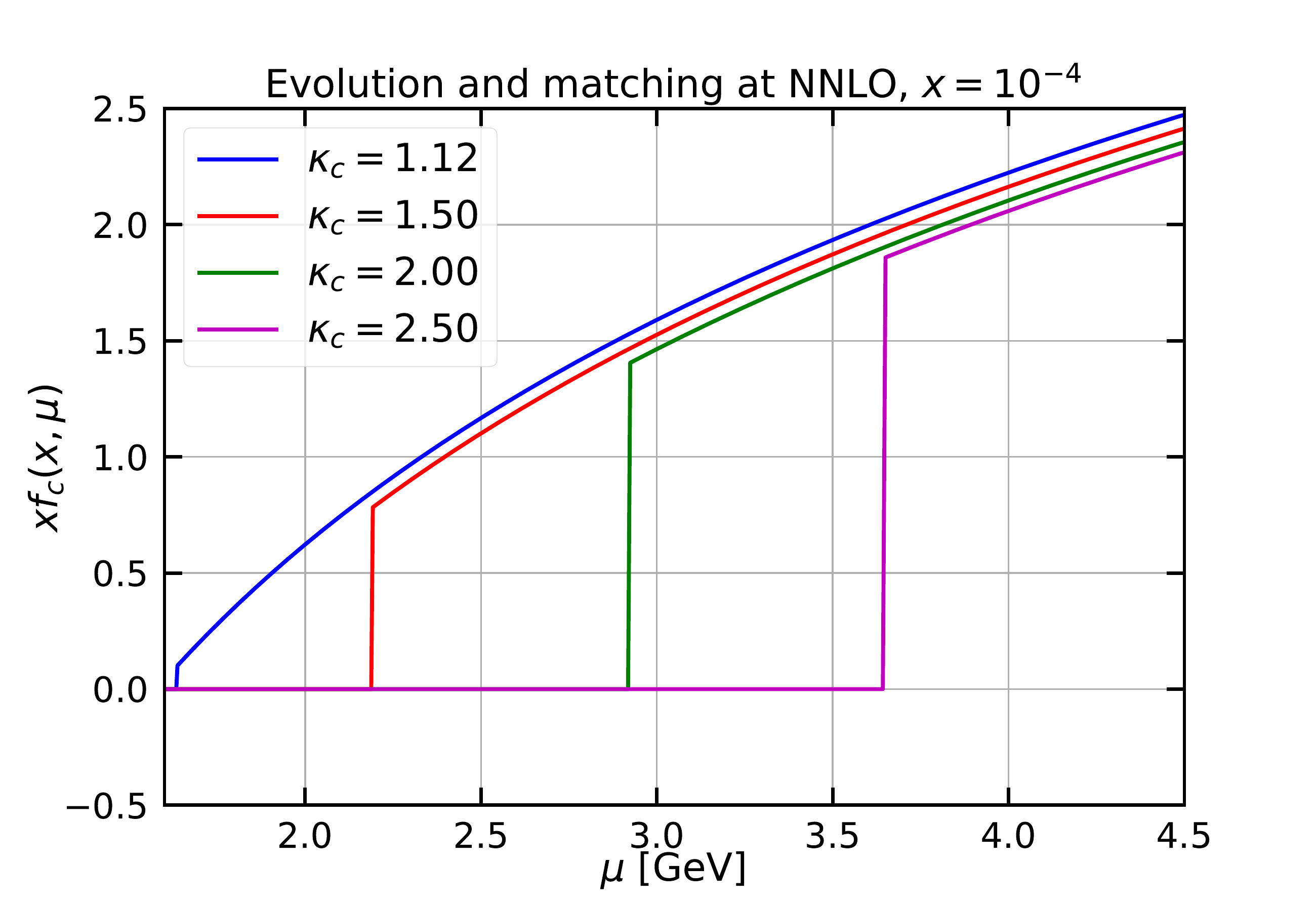}
    \includegraphics[width=0.49\textwidth]{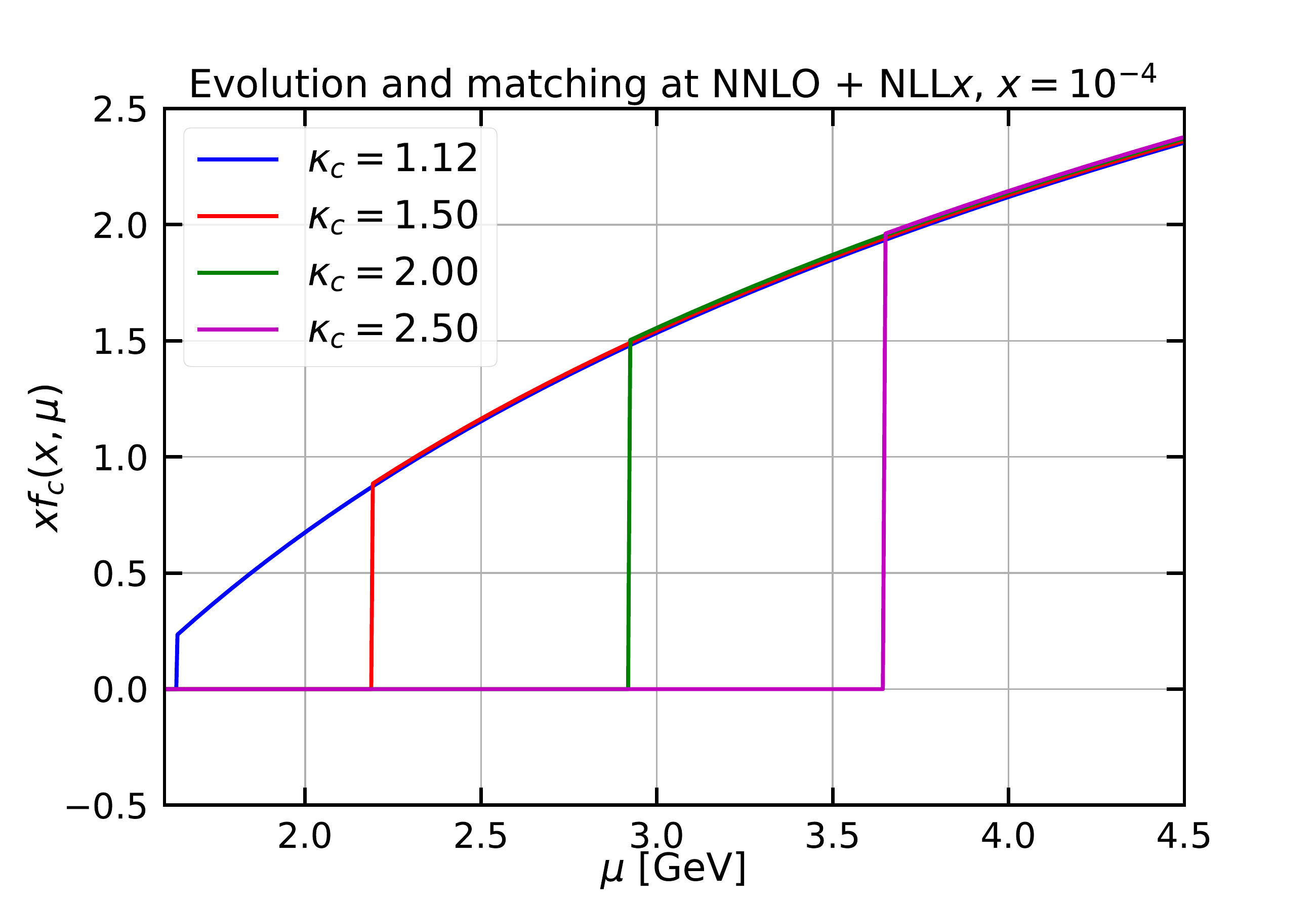}
    \caption{The charm PDF at $x=10^{-4}$ as a function of the
      factorisation scale $\mu$ for different values of the charm-quark
      matching scale $\mu_c = \kappa_c m_c$, with
      $\kappa_c=1.12,1.5,2,2.5$. The plots show the effect of the
      matching at NNLO (upper plot) and at NNLO+NLL$x$ (lower
      plot). \label{fig:matching}}
  \end{centering}
\end{figure}

In the next step the charm-quark matching scale $\mu_c$ is moved from
$\mu_c=m_c=1.43$~GeV to $\mu_c=1.12m_c\simeq1.6$~GeV.  The $\chi^2$
remains effectively unchanged, \textit{i.e.}\ $\chi^2=1390$.  Again,
PDFs do not change significantly, the only exception being the gluon
PDF which is slightly enhanced at low values of $x$.  The origin of
this enhancement can be traced back to the PDF matching
conditions. The upper plot of Fig.~\ref{fig:matching} shows that at
low $x$ ($x=10^{-4}$) moving up the charm-quark matching scale and
using fixed-order NNLO matching conditions has the effect of slightly
depressing the charm PDFs at large scales. Since in our fits charm is
generated dynamically mostly by gluon splitting, in order to describe
the charm component of the experimental data close to the charm-quark
matching scale, the gluon PDF must become slightly larger at low $x$
to compensate.  Interestingly, PDF matching conditions in this region
are also affected by large logarithms of $1/x$. These logarithms are
resummed in {\tt HELL} exactly like those in the splitting functions
and in the DIS coefficient functions.  Adding $\ln(1/x)$ resummation
to matching conditions and PDF evolution leads to the result in the
lower plot of Fig.~\ref{fig:matching}.  It is evident that the spread
caused by the matching (namely, the residual uncertainty from missing
higher orders) is significantly reduced when introducing
$\ln(1/x)$-resummation effects.  This shows that our results are
particularly stable upon displacement of the charm-quark matching
scale when resummation is included.

In the subsequent step the initial scale is then safely moved from
$Q_0^2=1.9$~GeV$^2$ to $Q_0^2=2.56$~GeV$^2$.  The $\chi^2$ does not
change significantly, {\it i.e.}\ $\chi^2=1388$.  Again, PDFs are
mostly unaffected with the exception of the gluon PDF, which is
enhanced, mostly at low values of $x$.  This change of the PDF shape
with $Q_0$ represents a parametrisation uncertainty. Such
uncertainties are included in the total PDF uncertainty and thus it
will be accounted for in the assessment of the impact of resummation
(see Fig.~\ref{fig:figdiff} later and discussion thereof).

Finally, in the last step $\ln(1/x)$ resummation at NLL$x$ is turned
on.  The $\chi^2$ of the fit falls to 1316. At this step the gluon PDF
at $Q^2=3$~GeV$^2$ differs significantly from that of HERAPDF2.0, and
also from that of the intermediate (fixed-order) steps, being much
steeper at low $x$ (see Fig.~\ref{fig:5steps}). The total singlet also
changes visibly at small $x$. Non-singlet quark PDFs, instead, are
insensitive to $\ln(1/x)$ resummation.

The above procedure clearly illustrates the improvement in $\chi^2$
deriving from $\ln(1/x)$ resummation, and the impact on the gluon and
singlet PDFs.  However, before assessing the significance of the
resummation by studying PDF uncertainties, a few refinements are still
needed.

Firstly, given that we now have a completely diffe\-rent shape of the
gluon PDF, it is necessary to investigate if the parametrisation used
for HERAPDF2.0 is adequate. A parametrisation scan was performed in
the FONLL-C scheme with $\ln(1/x)$ resummation and the parametrisation
of HERAPDF2.0 was confirmed. However, the negative term in the gluon
(see Eq.~(\ref{eqn:gluonpdf})) is now small, being compatible with
zero within three standard deviations, to be compared to more than
five standard deviations for HERAPDF2.0. In fact, this is also the
case for the fit without $\ln(1/x)$ resummation and is due to the
higher starting scale $Q^2_0=2.56$~GeV$^2$.  This suggests that the
parametrisation uncertainty previously found when raising $Q_0$ is
likely to be reduced at this scale.  Nevertheless, the shape of the
low-$x$ gluon is very different for the fits with and without
$\ln(1/x)$ resummation, being flattish/decreasing for the standard
NNLO fit and steep for the NNLO+NLL$x$ fit.  We return to this point
below.

\begin{figure*}[t]
  \begin{centering}
    \includegraphics[width=0.32\textwidth]{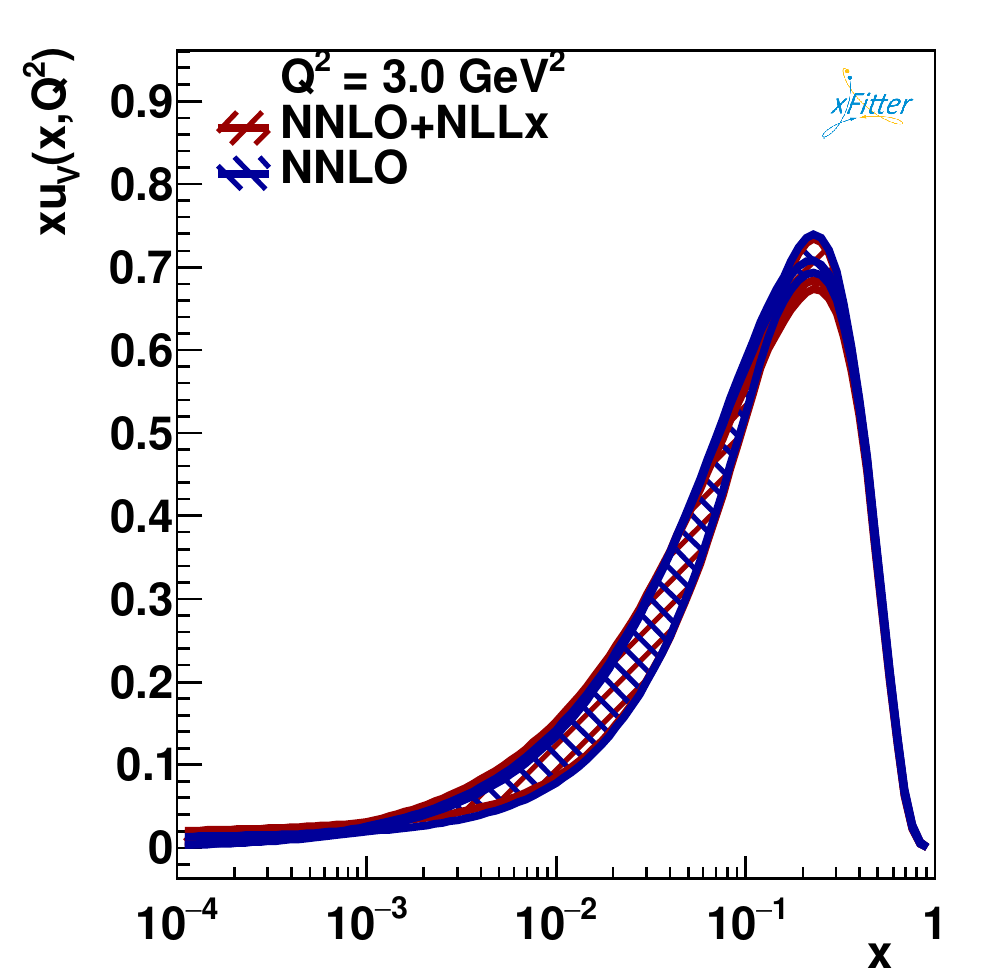}
    \includegraphics[width=0.32\textwidth]{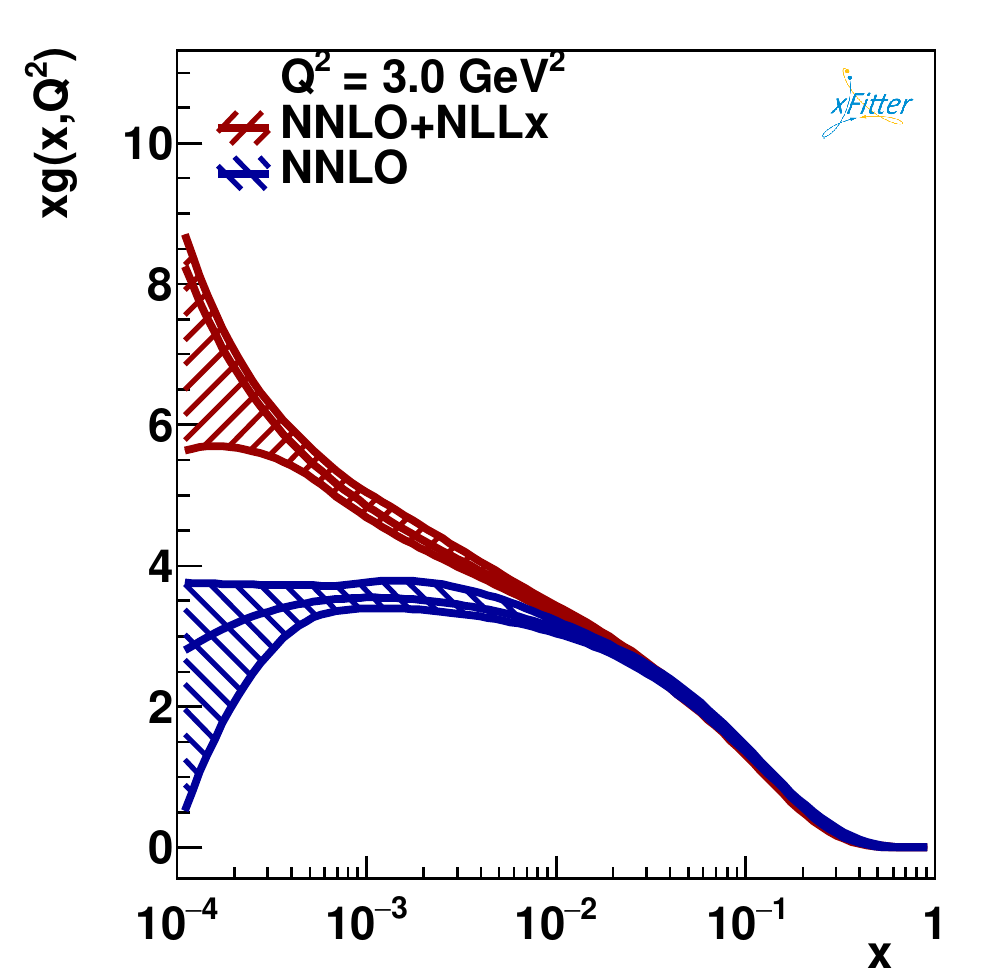}
    \includegraphics[width=0.32\textwidth]{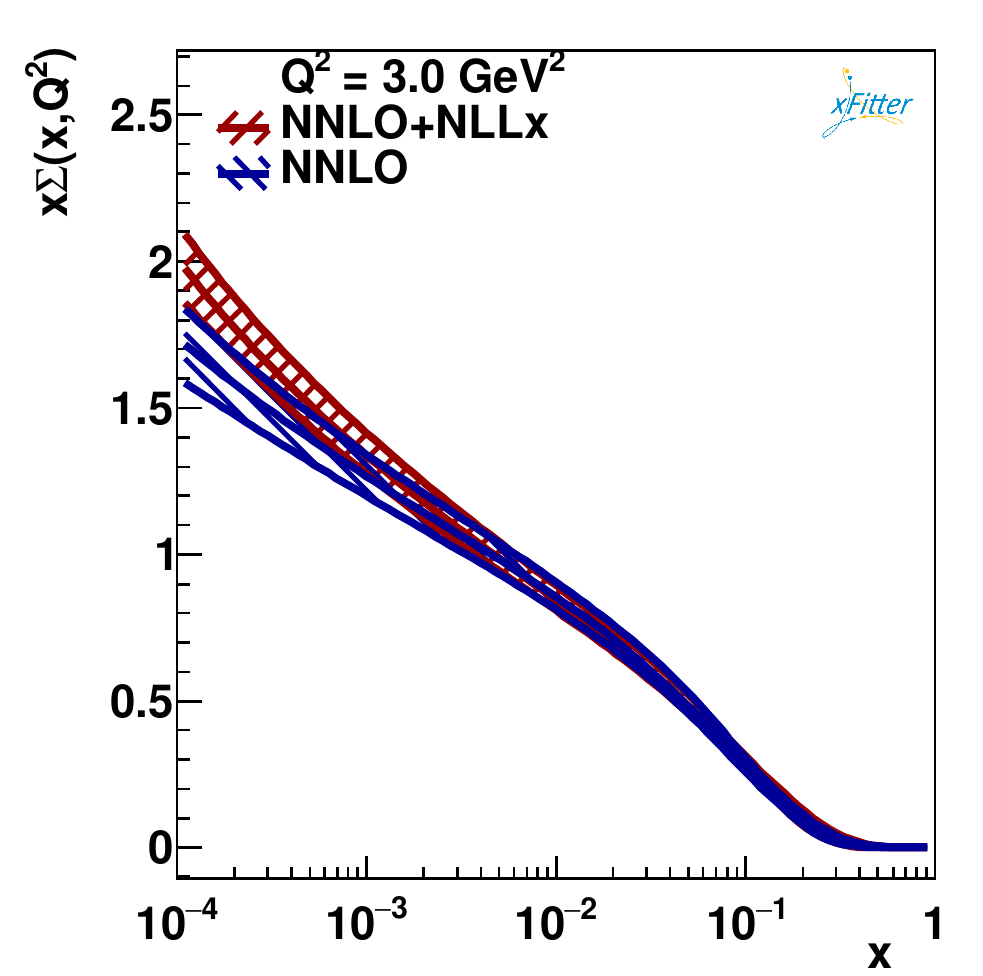}
    \caption{The up valence PDF $xu_v$, the gluon PDF $xg$ and the
      total singlet PDF $x\Sigma$ for the final fits with
      (NNLO+NLL$x$) and without (NNLO) $\ln(1/x)$ resummation.}
    \label{fig:finalpdf}
  \end{centering}
\end{figure*}
 
Secondly, the charm and beauty mass values used may not be optimal for
the new settings.  Thus charm~\cite{Abramowicz:1900rp} and beauty
data~\cite{Aaron:2009af,Abramowicz:2014zub} from HERA are included in
the fit to perform a charm-mass scan and a beauty-mass scan.  Various
fits have been performed by changing the charm and beauty pole masses
$m_c$ and $m_b$.  The optimal values $m_c=1.46$~GeV and $m_b=4.5$~GeV
are obtained from the NNLO+NLL$x$ fits.  The value of the charm-quark
matching scale is also moved accordingly to $\mu_c \simeq 1.64$~GeV,
to keep $\mu_c/m_c=1.12$.  Note that the $\chi^2$ variation over a
range of $0.1$~GeV for $m_c$ and $0.3$~GeV for $m_b$ reaches 1 unit at
most. This insensitivity is similar for both the fixed-order and
resummed fits, thus showing a good stability of the fits for the
variation of these physical parameters.

Since the charm data are in a kinematic region in which
$\ln(1/x)$-resummation corrections are important, this data set will
also be included in our final fits.  The beauty measurements mostly
lie at higher $x$ and $Q^2$ and thus are not expected to give a
significant contribution in the region of interest.  Indeed, we have
verified that including these data in the fit does not change the PDFs
in any appreciable way.  Moreover, the $\chi^2$ of the beauty datasets
computed from PDFs determined with and without those data are
basically the same.  This is mostly due to the fact that these
datasets contain only a very small number of datapoints (29 in total)
with large uncertainties.  While their inclusion in the fit does not
impact the PDF determination, we decided to retain them for our main
results.

\subsection{\textit{Final results with full uncertainties and comparison with data}}

The final fits that we are going to present use HERA inclusive, charm
and beauty data with the new values of $m_c=1.46$~GeV, $m_b=4.5$~GeV
and $Q^2_0=2.56$~GeV, and make use of the FONLL-C scheme, with and
without $\ln(1/x)$ resummation as implemented in {\tt HELL}.  An
exploration of various sources of the uncertainties has been
performed, following the HERAPDF2.0 prescription.  In addition to the
experimental uncertainty, which is evaluated using either the Hessian
(our default) or the Monte Carlo method, a number of model
uncertainties are considered.  Specifically, we have varied the charm
mass ($\Delta m_c=\pm 0.05$~GeV), the bottom mass
($\Delta m_b=\pm 0.25$~GeV), the strong coupling $\alpha_S(m_Z^2)$
($\Delta \alpha_S=\pm 0.001$), the fraction of strangeness
($\Delta f_s=\pm 0.1$), the initial scale ($Q^2_0=2.88$~GeV$^2$), and
the $Q_{\rm min}^2$ cut on the data ($Q^2_{\rm min}= 2.7$~GeV$^2$ and
$Q^2_{\rm min}=5$~GeV$^2$).  Additionally, parametrisation
uncertainties have been explored by adding extra terms to the
polynomials $P_i(x)$ of Eq.~\eqref{eqn:pdf}. This can give rise to
different PDF shapes with only slightly different $\chi^2$'s from that
of the main fit. In the present case, the only noticeable difference
comes from the inclusion of a linear term to the polynomial
$P_{u_v}(x)$ of the valence up quark PDF (this was also found in the
HERAPDF2.0 study).  The largest difference on the uncertainty of the
gluon distribution arises from the variation of the $Q^2_{\rm min}$
cut to $5$~GeV$^2$. Interestingly, this uncertainty decreases for the
fit with $\ln(1/x)$ resummation due to the reduced tensions with the
data, see the discussion below.

\begin{figure}[t]
\begin{center}
  \includegraphics[width=0.4\textwidth]{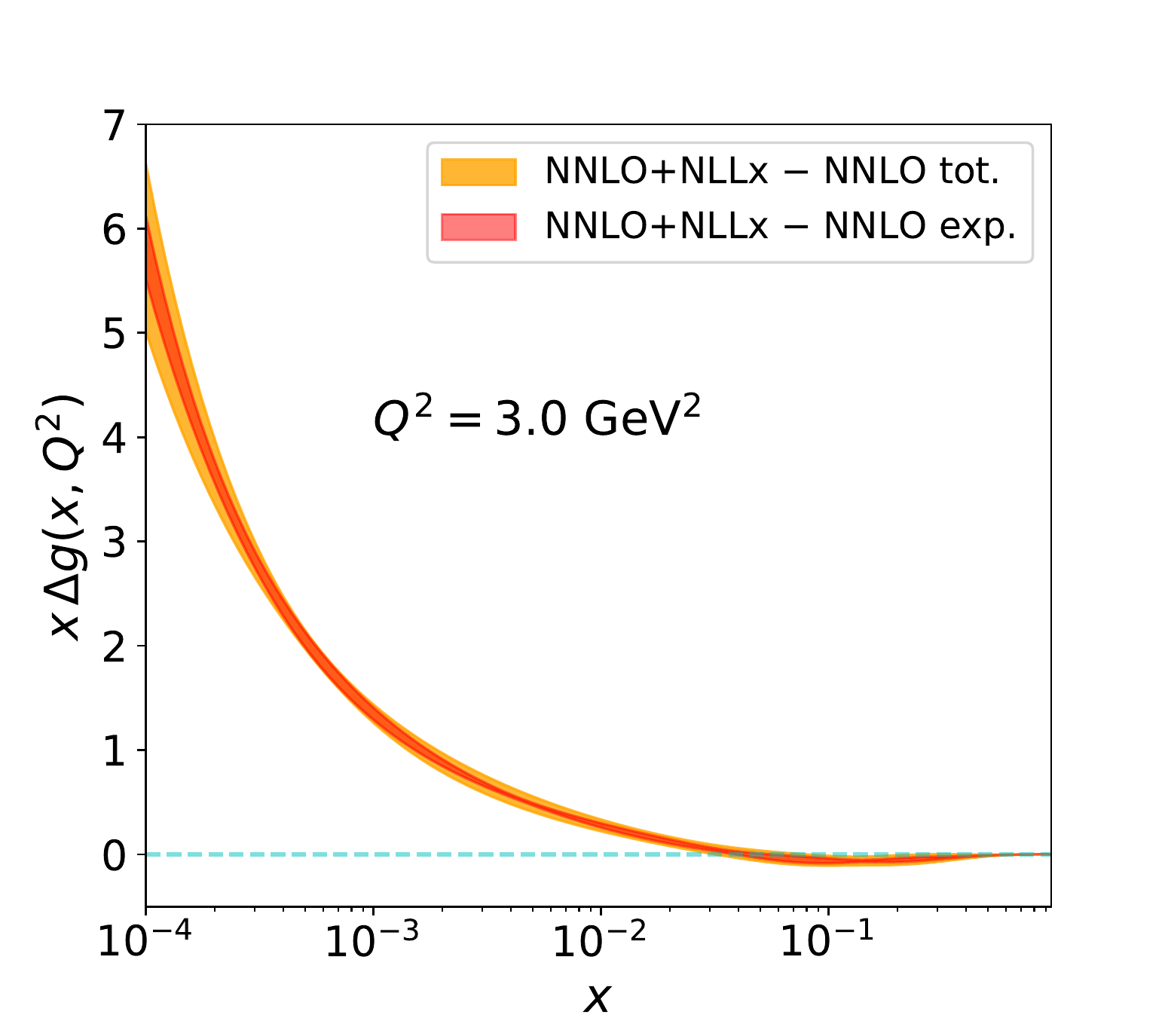}
  \caption{The difference between the gluon distribution determined in
    the fits at NNLO with and without NLLx resummation taking into
    account the correlations between their uncertainties. The orange
    (red) band indicates the full (experimental) uncertainty on the
    difference.}
\label{fig:figdiff}
\end{center}
\end{figure}

\begin{figure*}[t]
\begin{center}
  \includegraphics[width=0.65\textwidth,trim=2.1cm 0 2.1cm 0]{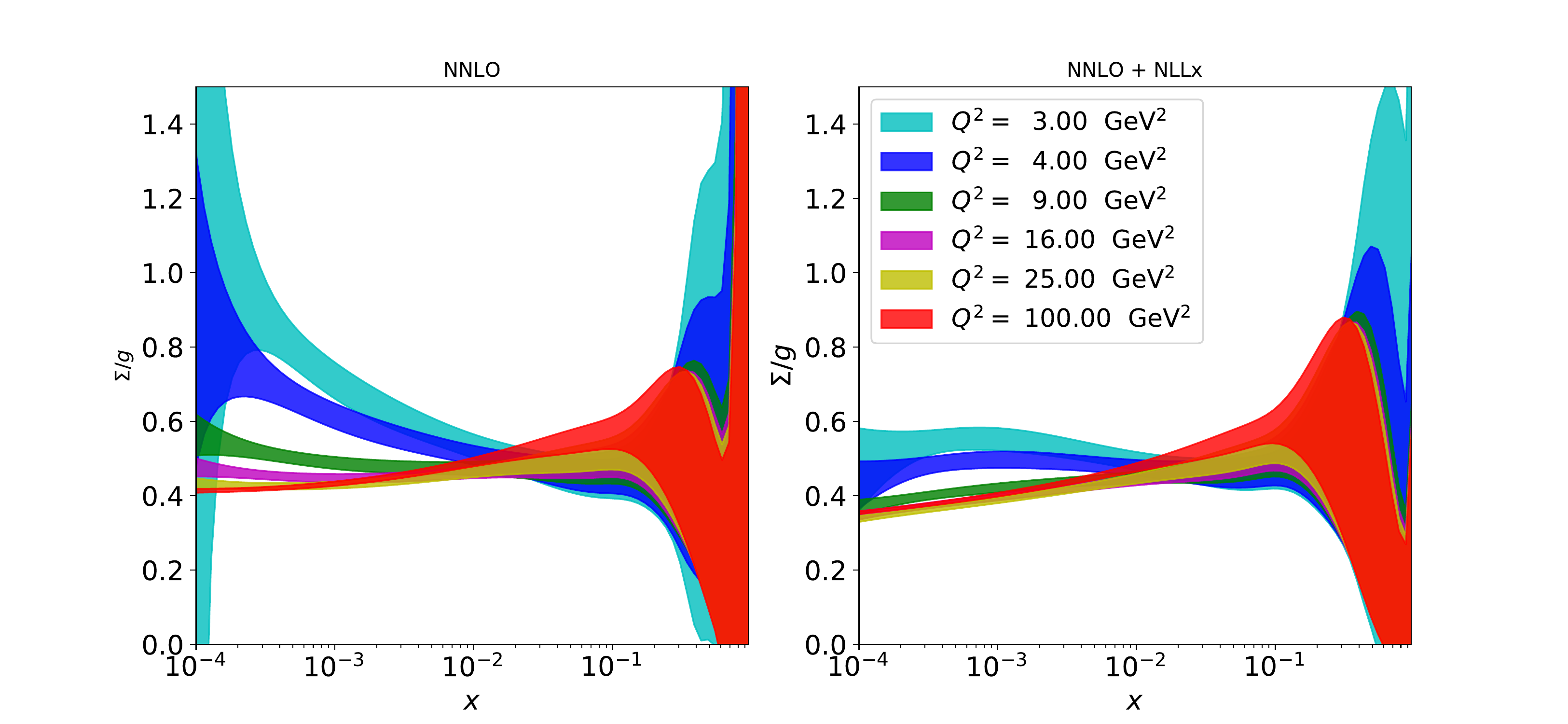}
  \caption{The ratio of the total singlet to the gluon PDF as a
    function of $x$ shown for different scales $Q^2$ for the final
    fits with (right plot) and without (left plot) $\ln(1/x)$
    resummation.}
  \label{fig:sigglu}
\end{center}
\end{figure*}

Fig.~\ref{fig:finalpdf} shows a direct comparison of PDFs with and
without $\ln(1/x)$ resummation at $Q^2 = 3$~GeV$^2$.  This figure
displays also the full uncertainty bands.  Note, however, that since
the data used in the two fits are the same, the uncertainty bands are
highly correlated. In order to quantify the difference in the gluon
shape taking into account the correlations, the method developed in
Ref.~\cite{Belov:2014xwo} is employed.  Specifically, we take a
version of our fits with experimental uncertainties estimated using
the Monte Carlo method. We then generate replicas of the data using
the same random number sequence used for the fits with and without
resummation to evaluate the spread of the synchronised differences. A
si\-mi\-lar procedure is adopted for the model and parameterisation
uncertainties (including the uncertainty due to variation of $Q_0$
mentioned above). The difference for the gluon distribution with its
uncertainty is shown in Fig.~\ref{fig:figdiff}. The correlated PDF
sets at NNLO and NNLO+NLLx can be used to evaluate the impact of
$\ln(1/x)$ resummation on other analyses and can be found on the {\tt
  xFitter} public page.

\begin{figure}[t]
  \centering
  \includegraphics[width=0.49\textwidth]{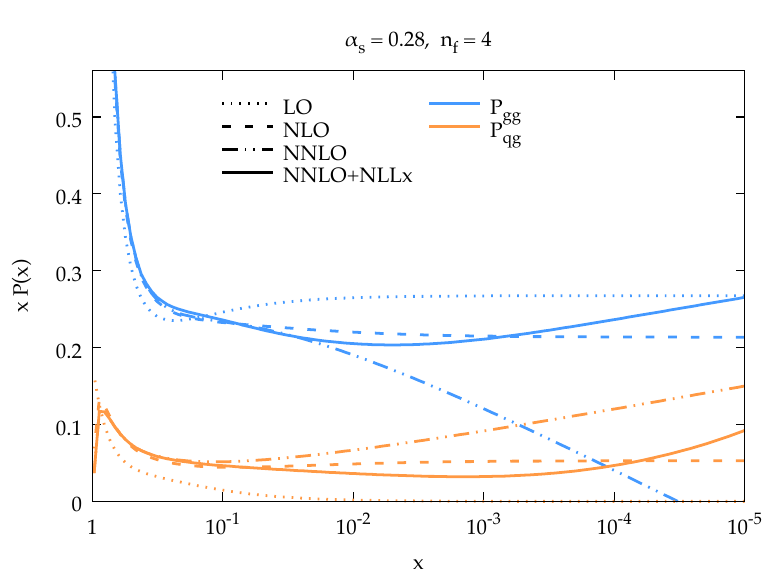}
  \caption{The resummed splitting functions at NNLO+NLL$x$ (solid)
    compared to fixed order at LO (dotted), NLO (dashed) and NNLO
    (dot-dot-dashed) for $P_{gg}$ (upper curves) and $P_{qg}$ (lower
    curves) as a function of $x$. The plots are at
    $\alpha_S(Q^2)=0.28$ (corresponding to $Q^2\sim 4$~GeV$^2$) and
    $n_f=4$.}
  \label{fig:split}
\end{figure}

When resummation is included, both the gluon and the total singlet
PDFs rise at low $x$. This contrasts with the shape of the gluon when
resummation is not included. This behaviour can be studied in more
detail by examining the evolution of the ratio $\Sigma/g$ at different
scales, as shown in Fig.~\ref{fig:sigglu}.  For the fit without
resummation, the ratio exhibits a strong dependence on the scale,
ranging from values exceeding unity at low $x$ and low scales to
values $\sim0.5$ at high scales. The evolution of the ratio is much
more stable when resummation is included, with the total singlet never
exceeding the gluon PDF and remaining approximately constant at low
$x$. At large scales ($Q\sim 1000$~GeV) and low $x$, the ratio
$\Sigma/g$ becomes equal to within a few percent for the fits with and
without resummation, while gluon and total singlet remain different at
the $50\%$ level. This behaviour of the ratio for the fit without
resummation can be explained by a peculiar feature of the $xP_{gg}(x)$
and $xP_{qg}(x)$ splitting functions, see Fig.~\ref{fig:split}.
For $Q^2\sim 4$~GeV$^2$ ($\alpha_S(Q^2) =0.28$), the NNLO splitting
function $xP_{gg}(x)$ ($xP_{qg}(x)$) falls (rises) for $x\to 0$ with
$xP_{qg}(x)>xP_{gg}(x)$ for $x \lesssim 10^{-3}$. This causes a
suppression of the low-$x$ gluon in favour of the total singlet
PDF. When resummation is added, the relation $xP_{qg}(x)<xP_{gg}(x)$
is restored down to very low values of $x$ leading to a suppression of
the total singlet compared to the gluon PDF.

\begin{table*}
\begin{center}
\begin{tabular}{lcc}
  \toprule
  &NNLO fit &  NNLO+NLL$x$ fit  \\
  &  with new settings      &  with new settings  \\
  \midrule
  Total $\chi^2 (=\tilde\chi^2+\text{corr}+\log)/\rm{d.o.f.}$       & $1468\,(1327+119+22)/1207$ & $1394\,(1305+91-2)/1207$ \\
  \midrule
  dataset inclusive $(\tilde\chi^2+\text{corr}+\log)/\rm{n.d.p.}$   & $(1264+103+21)/1145$   & $(1239+78-4)/1145$   \\
  \quad - subset NC 920    $\tilde\chi^2/\rm{n.d.p.}$   & $447/377$   & $413/377$   \\
  \quad - subset NC 820    $\tilde\chi^2/\rm{n.d.p.}$   & $67/70$     & $65/70$   \\
  dataset charm     $(\tilde\chi^2+\text{corr}+\log)/\rm{n.d.p.}$   & $(47+12-1)/47$     & $(50+11-1)/47$  \\
  dataset beauty    $(\tilde\chi^2+\text{corr}+\log)/\rm{n.d.p.}$   & $(16+2+3)/29$     & $(16+2+3)/29$  \\
  \bottomrule
\end{tabular}
\caption{Total $\chi^2$ per d.o.f. and some of the partial
  $\tilde\chi^2$'s per number of data points (n.d.p.) for the PDF fits
  to HERA inclusive and heavy-quark data with and without $\ln(1/x)$
  resummation with the new settings. Also shown are the contributions
  to the $\chi^2$ from the correlated shifts and the log terms.}
\label{tab:fitresults2}
\end{center}
\end{table*}

The $\chi^2$ values of the final fits are summarised in
Tab.~\ref{tab:fitresults2}. There is a decrease of $74$ units in
$\chi^2$ when $\ln(1/x)$ resummation is used. Most of this difference
comes from the highly accurate NC $E_p = 920$~GeV data which probe the
low-$x$ and low-$Q^2$ region and thus are particularly sensitive to
$\ln(1/x)$ resummation. The table also shows the partial $\chi^2$ for
these data, the NC $E_p=820$~GeV\footnote{In the HERAPDF2.0
  combination, the data at $E_p=820$~GeV and $y<0.35$ are combined
  with $E_p=920$~GeV and attributed to the $E_p=920$~GeV data
  set. Thus $E_p=820$~GeV data contain only the high $y>0.35$ region
  which enhances sensitivity to the $\ln(1/x)$ resummation effects.}
and charm and beauty data, which may also be sensitive.  Other data
sets entering the fit probe higher $x$ and $Q^2$ and their $\chi^2$ is
not significantly affected, and so they are not shown in the table.
To fully appreciate the source of the overall improvement in $\chi^2$,
it is necessary to consider the contribution due to the correlated
systematic uncertainties and the lo\-ga\-rith\-mic term. The form of
the $\chi^2$ minimised during the fits is given
by~\cite{Aaron:2012qi}:
\begin{equation}\label{eq:chi2}
\begin{array}{rcl}
  \chi^2&=&\displaystyle \sum_i\frac{\left[ D_i -T_i\left(1-\sum_j\gamma_j^i
  b_j\right) \right]^2}{\delta_{i,{\rm unc}}^2 T_i^2+\delta_{i,{\rm
  stat}}^2D_i T_i} + \sum_j b_j^2\\
\\
&+&\displaystyle \sum_i \ln \frac{\delta_{i,{\rm
  unc}}^2 T_i^2+\delta_{i,{\rm stat}}^2D_i T_i}{\delta_{i,{\rm unc}}^2
  D_i^2+\delta_{i,{\rm stat}}^2D_i^2},
\end{array}
\end{equation}
where $T_i$ is the theoretical prediction and $D_i$ the measured value
of the $i$-th data point, $\delta_{i,{\rm stat}}$,
$\delta_{i,{\rm unc}}$, and $\gamma_j^i$ are the relative statistical,
uncorrelated systematic, and correlated systematic uncertainties, and
$b_j$ are the nuisance parameters associated to the correlated
systematics which are determined during the fit.  The
``$\tilde\chi^2$'', ``corr'' and ``log'' contributions reported in
Tab.~\ref{tab:fitresults2} correspond to the first, second and third
terms in the r.h.s. of Eq.~\eqref{eq:chi2}, respectively. A reduction
of the correlated shifts term indicates that the fit does not require
the predictions to be shifted so far within the tolerance of the
correlated systematic uncertainties, while a reduction of the log term
reflects a better agreement of the theoretical predictions with the
data. Considering the partial $\tilde\chi^2$, the correlated shift
term and the log term for the inclusive and heavy-quark data, we can
see that the largest improvement comes from the NC $E_p=920$~GeV,
which is much better described in the NNLO+NLLx fit.  There is no
visible improvement in the NC $E_p=820$~GeV data set, perhaps due to
the larger uncertainties of its low $x$ data.  There is no improvement
for the beauty data either, and since most of the data points are at
higher $x$ and $Q^2$ this is not surprising.  More surprisingly, the
change for the charm data from $\chi^2=58$ at NNLO to $\chi^2=60$ at
NNLO+NLLx is negligible.  This contrasts with the results of
Ref.~\cite{Ball:2017otu}. The origin of this diffe\-ren\-ce is that
the FONLL scheme with perturbatively ge\-ne\-ra\-ted charm at NNLO,
used in this analysis, provides a better description of the charm data
than the FONLL im\-ple\-men\-ta\-tion with fitted
charm~\cite{Ball:2015tna,Ball:2015dpa} (as also found in
Ref.~\cite{Ball:2017otu}).  We will return on this at the end of the
section.

\begin{figure}[t]
  \begin{centering}
    \includegraphics[width=0.49\textwidth]{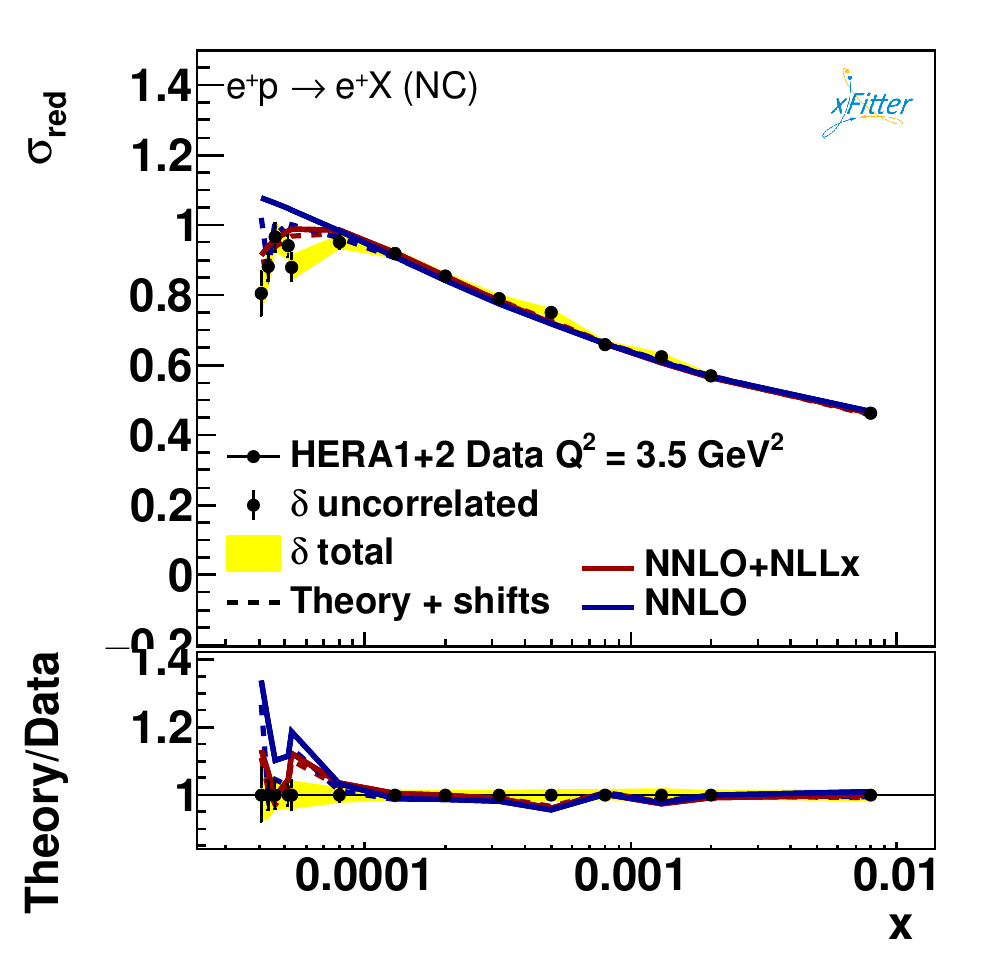}
    \includegraphics[width=0.49\textwidth]{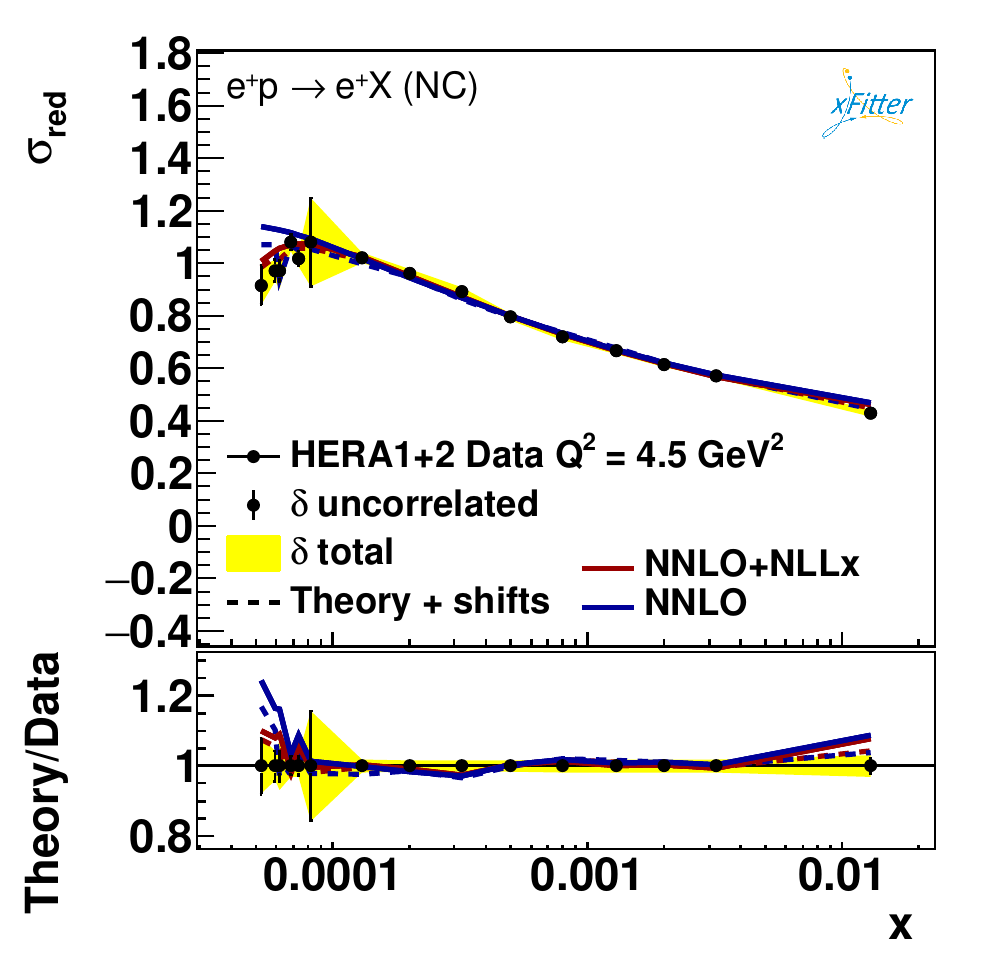}
    \caption{The HERA NC $E_p= 920$~GeV data compared to the fits with
      and without $\ln(1/x)$ resummation for the $Q^2 = 3.5$ and
      $4.5$~GeV$^2$ bins.\label{fig:data920}}
  \end{centering}
\end{figure} 
In Fig.~\ref{fig:data920} the results of the fits are compared to the
NC $E_p=920$~GeV inclusive reduced cross-section data in the lowest
$Q^2$ bins included in the fits. The plots illustrate the predictions
both before and after the shifts due to the experimental correlated
systematics are applied. The shift to the theoretical prediction
$T_i$, according to Eq.~(\ref{eq:chi2}), is given by
$T_i\sum_j\gamma_j^i b_j$. It is evident that for the fit including
$\ln(1/x)$-resummation effects the initial description of the data is
better and thus the correlated shifts are smaller. In particular, the
low-$x$ turn-over of the measurements is better reproduced by the fit
that includes $\ln(1/x)$ resummation. This is a direct consequence of
the steeper gluon at low $x$ (see Fig.~\ref{fig:finalpdf}) that makes
$F_L$ larger at low $x$ causing a more pronounced turn-over of the
reduced cross section (\textit{cfr.}  Eq.~(\ref{eq:redxsec})). This is
the main reason for the reduction in $\chi^2$ of the fit with
$\ln(1/x)$ resummation.

\begin{figure}[t]
  \begin{centering}
    \includegraphics[width=0.49\textwidth]{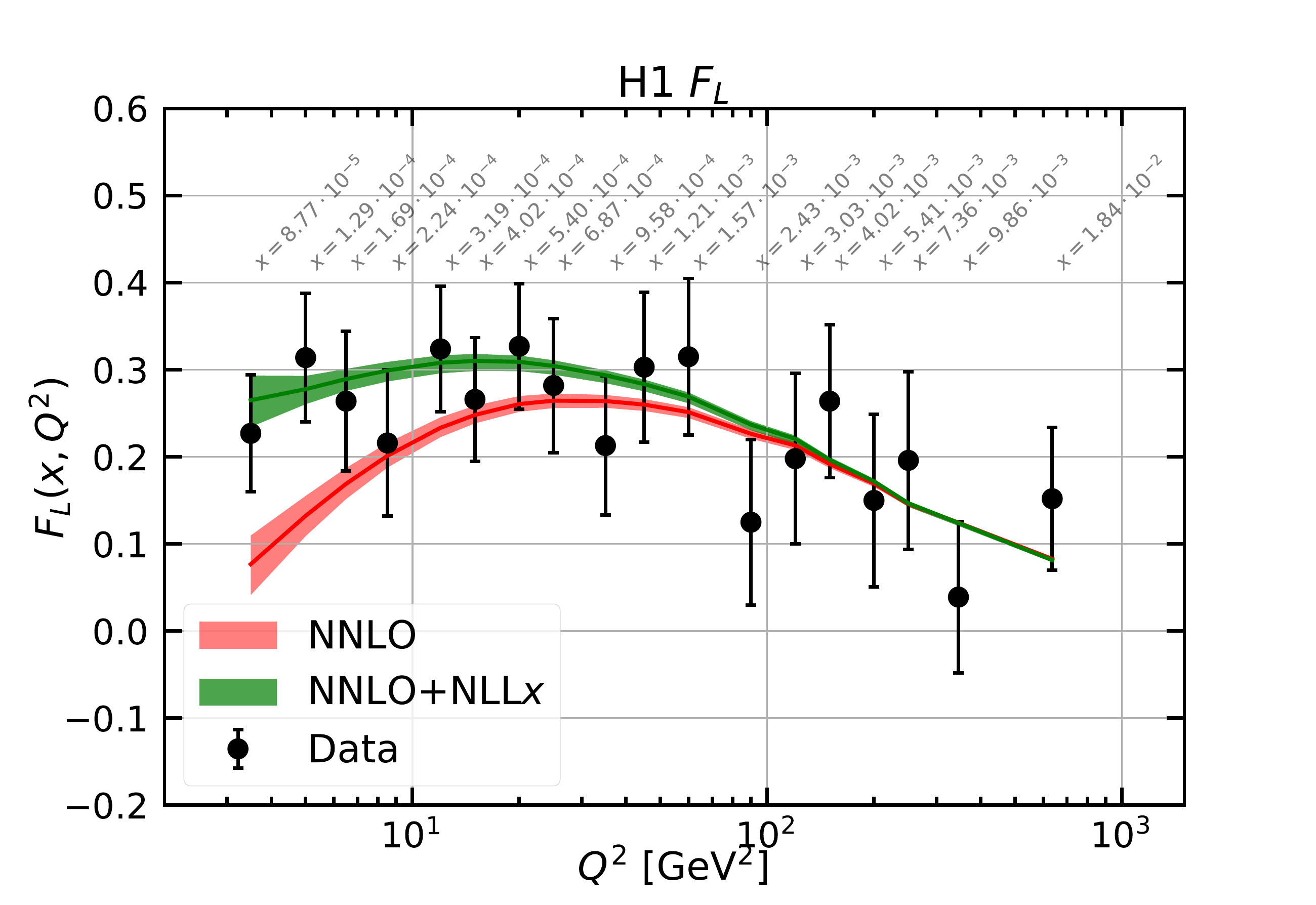}
    \caption{The H1 extraction of $F_L$ compared to the predictions
      with and without $\ln(1/x)$ resummation.\label{fig:f2fl}}
  \end{centering}
\end{figure}
This point is illustrated also in Fig.~\ref{fig:f2fl} where the
theo\-re\-ti\-cal predictions of $F_L$ with and without $\ln(1/x)$
resummation are compared to the H1 $F_L$ extraction. The visual
description of this data set is improved in the former case thanks to
the fact that $\ln(1/x)$-resummed predictions for $F_L$ are larger at
low $x$.\footnote{We recall that these data are \emph{not} explicitly
  included in our fit, but information on $F_L$ is included in the
  reduced cross sections which are fitted.}

\subsection{\textit{Comparison with the NNPDF analysis}}
\label{sec:nnpdf}

\begin{figure*}
  \centering
  \includegraphics[width=0.33\textwidth]{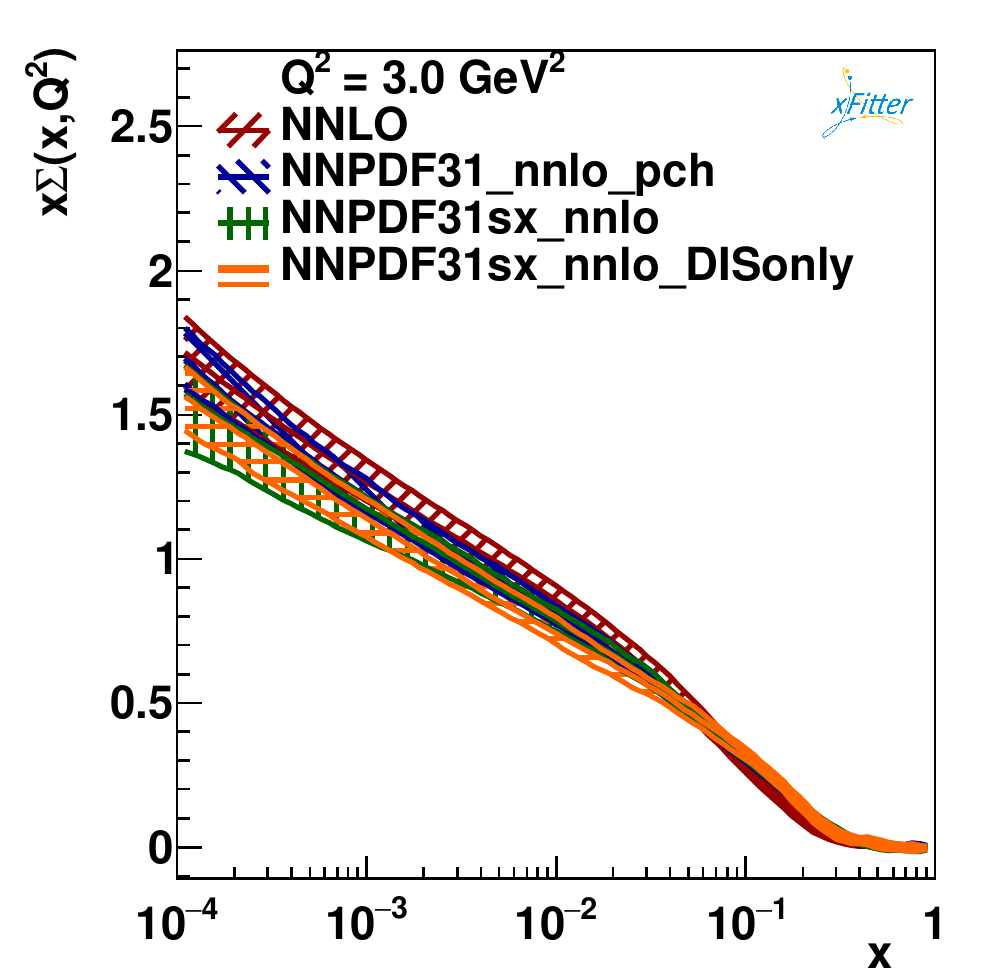}
  \includegraphics[width=0.33\textwidth]{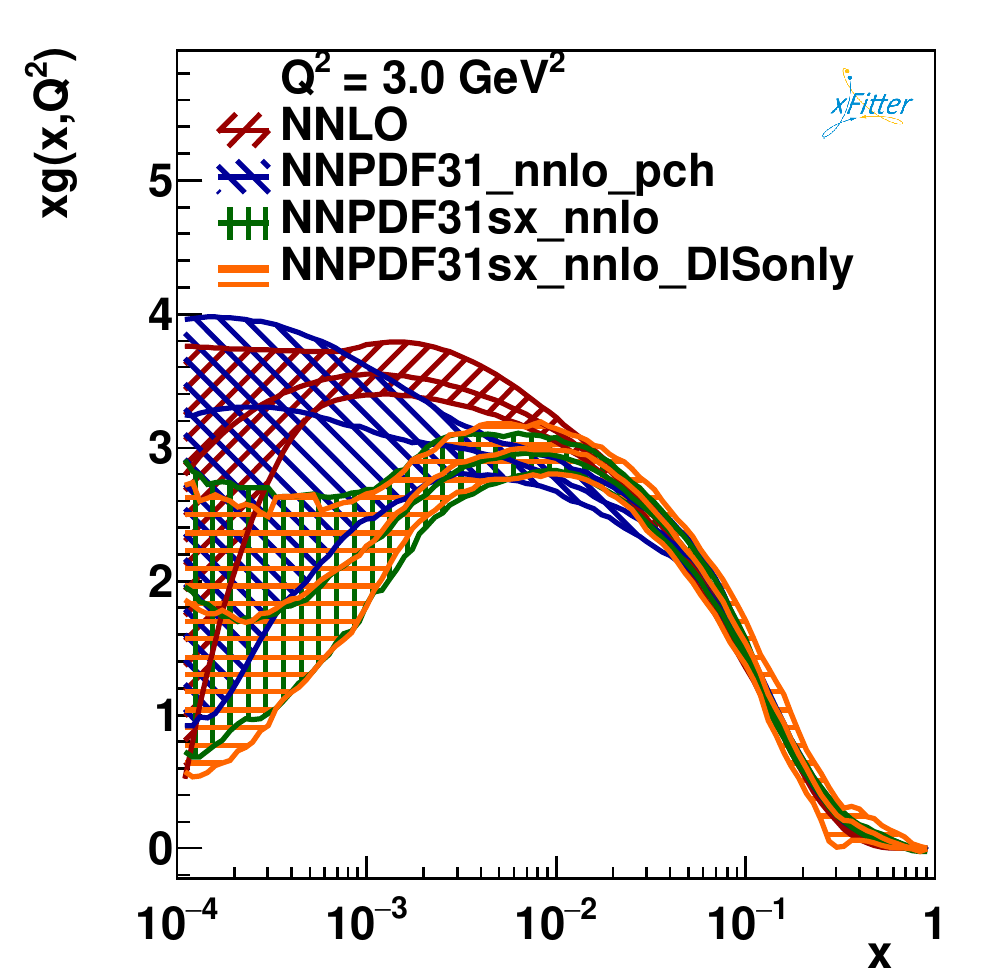}
  \includegraphics[width=0.33\textwidth]{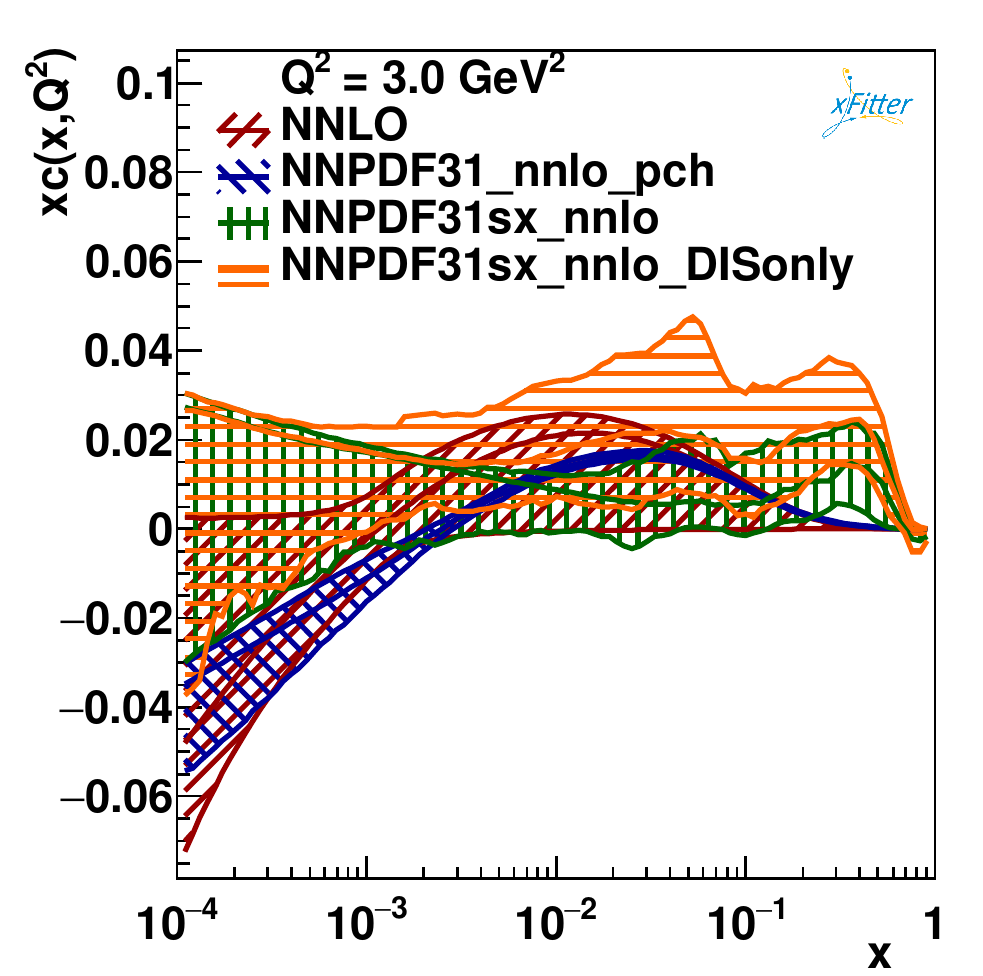}
  \includegraphics[width=0.33\textwidth]{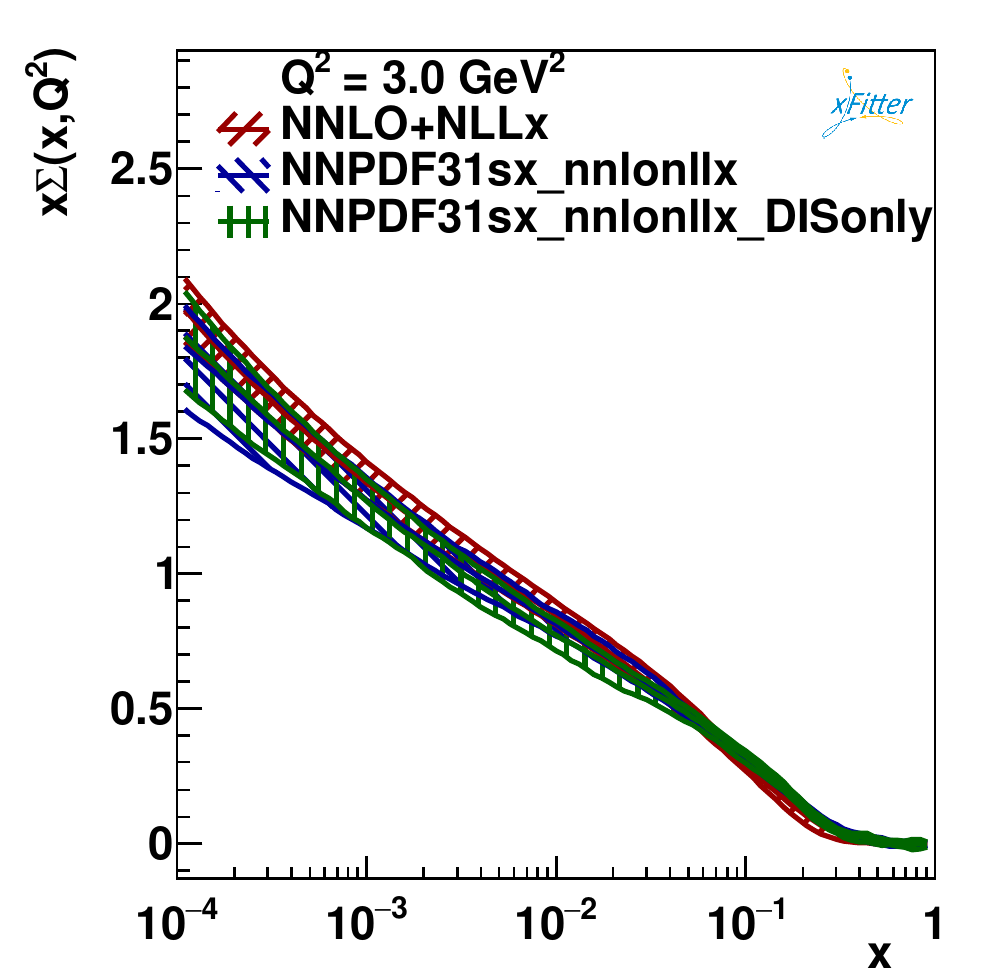}
  \includegraphics[width=0.33\textwidth]{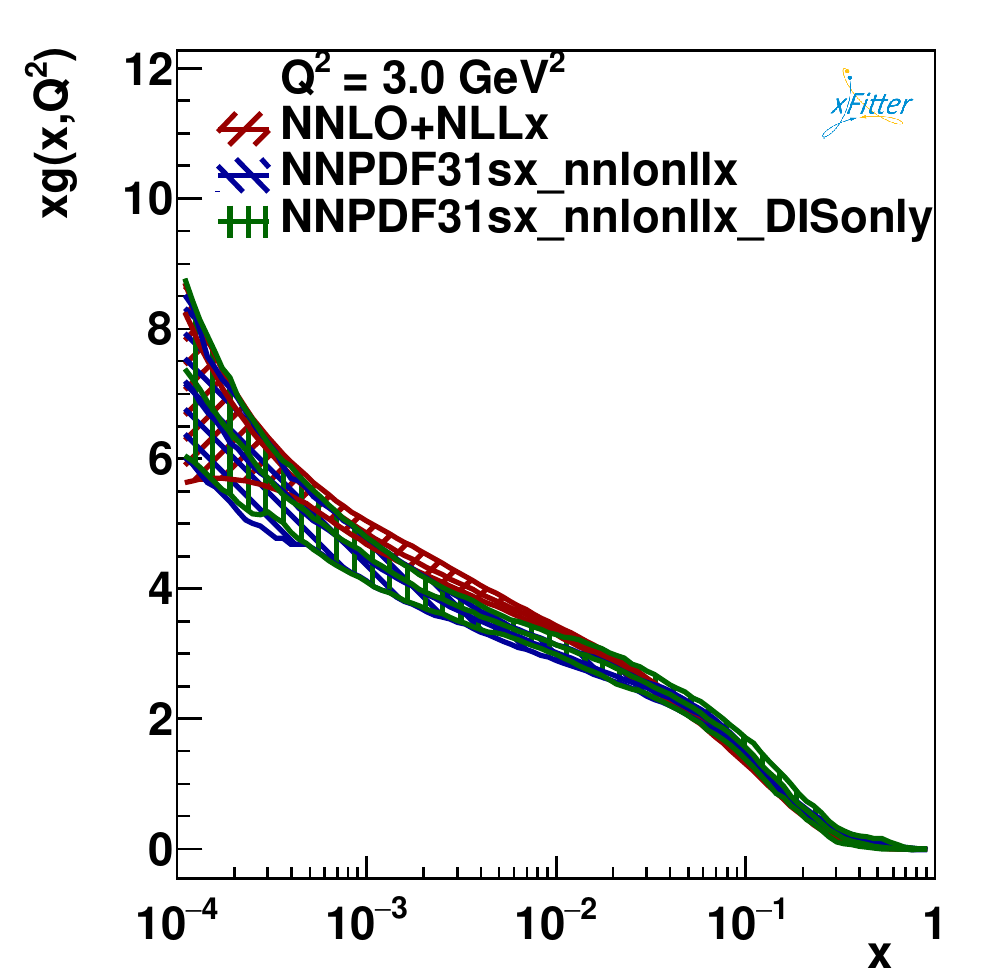}
  \includegraphics[width=0.33\textwidth]{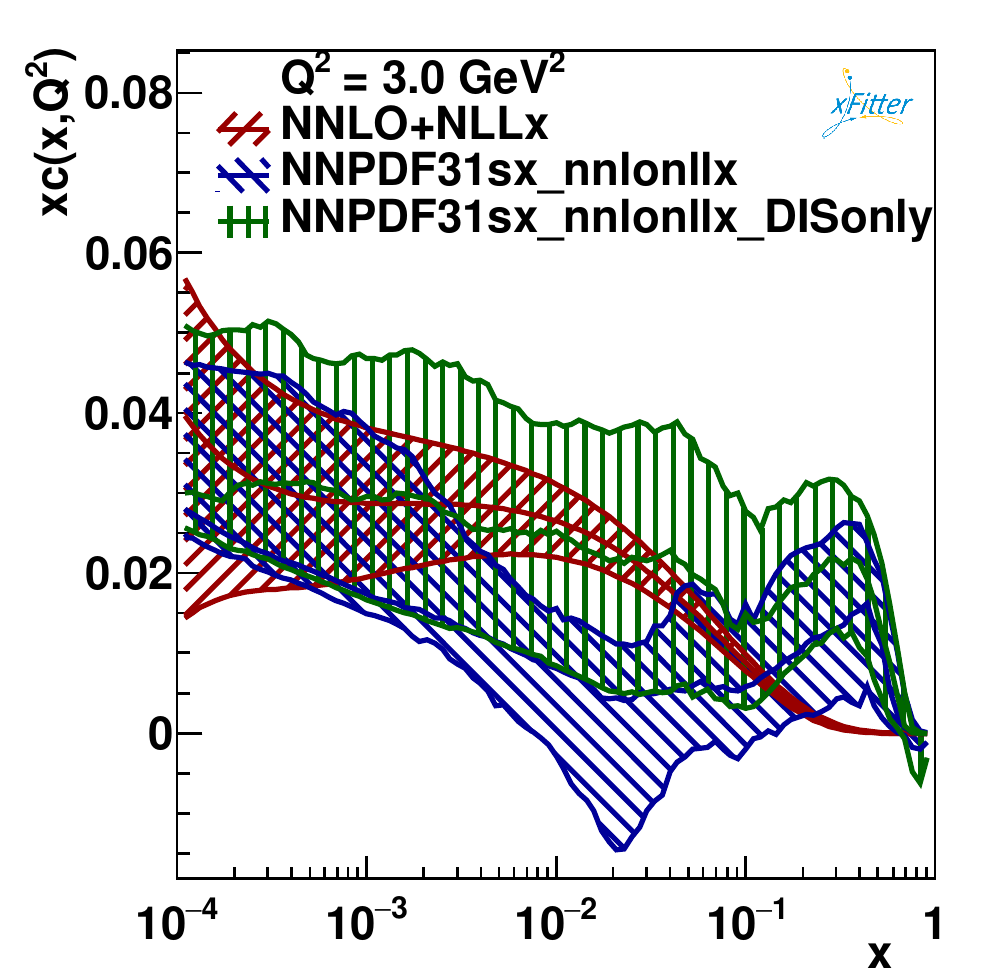}
  \caption{The total singlet, gluon and charm PDFs for the final fits
    at NNLO (upper plots) and NNLO+NLL$x$ (lower plots) compared to
    the analogous NNPDF3.1 determinations.}
  \label{fig:glunnpdf31}
\end{figure*}
We conclude this section by comparing our results with those of the
NNPDF3.1 family~\cite{Ball:2017otu, Ball:2017nwa}.  In
Fig.~\ref{fig:glunnpdf31} we show the total singlet, gluon and charm
PDFs with (lower plots) and without (upper plots) $\ln(1/x)$
resummation.  In particular, on top of our PDFs, we consider the
global and the DIS-only PDF sets of Ref.~\cite{Ball:2017otu}
(NNPDF3.1sx, henceforth). In contrast with this analysis, all
NNPDF3.1sx sets have been obtained by fitting the charm PDFs to data.
Therefore, in order to gauge the impact of the different treatments of
the charm PDFs, in the comparisons at fixed order we also consider the
NNPDF3.1 set at NNLO of Ref.~\cite{Ball:2017nwa} obtained using
perturbative charm.

At fixed order (upper plots of Fig.~\ref{fig:glunnpdf31}), the
quark-singlet PDF (left plot) appears to be very similar for all four
PDF sets considered. The gluon PDF (central plot), instead, presents
larger discrepancies. In particular, the NNPDF3.1sx distributions
(both global and DIS-only) are somewhat different from the gluon
obtained in this analysis at small $x$. Given the consistency of the
NNPDF3.1sx results, this appears to be the consequence of the
different treatment of the charm PDFs rather than the different data
sets. The gluon PDF of the NNPDF3.1 set with perturbative charm is
closer to our result at low-$x$ ($x \lesssim 0.001$) than to the
NNPDF3.1sx curves. We also observe that the suppression on the gluon
PDF of the NNPDF3.1sx sets causes an enhancement of the charm
distribution at small $x$ as compared to both this analysis and
NNPDF3.1 with perturbative charm (right plot).

Moving to the PDFs with $\ln(1/x)$ resummation (lower plots of
Fig.~\ref{fig:glunnpdf31}), we observe a better agreement between all
PDF sets considered. Noticeably, the gluon distributions are now
compatible despite the different treatment of the charm. As a
consequence, also the charm PDFs at small $x$ are in much better
agreement.  Note also that the uncertainty bands obtained in this
analysis are comparable to those of the NNPDF sets, except for the
charm PDF at large $x$ whose band is larger for the NNPDF3.1sx sets
due to the fact that the charm PDFs are fitted to data.

Another striking difference with respect to our analysis is that a
significant reduction (by more than 50 units for 47 datapoints) of the
$\chi^2$ of charm production data when including $\ln(1/x)$
resummation was found in Ref.~\cite{Ball:2017otu}.  The origin of such
a huge effect can be traced back to the poor quality of the
description of charm data at fixed NNLO in the NNPDF3.1sx
study. Indeed, the NNPDF3.1sx $\chi^2$ of this dataset when
resummation is included is very similar to that of the present study,
differing by just 2 units.  The reason of this difference in the
quality of the description of charm data at fixed order is related to
the treatment of the charm PDFs.  However, the discrepancy cannot be
ascribed to the fact that the charm PDFs are fitted in
Ref.~\cite{Ball:2017otu}. In fact, fitting the charm PDFs should give
more flexibility to better describe the data. Rather, it is the actual
construction of the FONLL-C prediction which differs when the charm
PDFs are fitted.  Specifically, when fitting the charm PDFs, it has
been pointed out that an extra contribution, denoted by $\DIC$, should
be added to the FONLL formula to account for potential
intrinsic-charm-initiated
contributions~\cite{Ball:2015tna,Ball:2015dpa}.

The introduction of this extra term has the consequence that the
phenomenological damping factor usually introduced in the FONLL scheme
with perturbative charm to suppress subleading higher-order terms in
the vicinity of the charm threshold~\cite{Forte:2010ta}, becomes
ineffective. Indeed, when the charm PDFs are fitted, and thus a
non-perturbative (or intrinsic) component is allowed, the
contributions multiplied by the damping are no longer subleading, and
cannot therefore be suppressed.
The bad description of the charm data at fixed order in
Ref.~\cite{Ball:2017otu} is thus the consequence of three concurring
effects: (1) the absence of damping, (2) the presence of the extra
contribution $\DIC$ to the FONLL formula, and (3) the fitted charm
PDFs which makes this $\DIC$ contribution sizeable. Since our charm
PDFs are generated perturbatively, the $\DIC$ contribution is
subleading and does not affect our results significantly.
Specifically, the effect of adding such $\DIC$ term would effectively
be equivalent to removing the damping factor.  We have thus performed
a fixed-order fit without the damping in the FONLL formula and found
that, as expected, the results are not significantly affected (in
particular, the $\chi^2$ of the charm dataset remains unchanged).  We
have also recomputed the $\chi^2$ of the charm dataset using FONLL
without damping and the NNLO PDFs of Ref.~\cite{Ball:2017otu}, which
contain fitted charm, and found indeed that the description of the
data is worsened, even though not at the level of the results of
Ref.~\cite{Ball:2017otu} (which additionally include the extra $\DIC$
contribution to FONLL).  Note that the deterioration of $\chi^2$ in
this case comes mostly from the correlated contribution to the
$\chi^2$, second term in Eq.~\eqref{eq:chi2}.  We have also performed
the same exercise activating the damping, which effectively suppresses
all contributions due to the fitted-charm PDFs in the vicinity of the
charm threshold making the result closer to what one obtains in the
perturbative charm case. By doing so we find that the description
improves significantly, bringing it at the level of our results.  Note
that similar tests have been performed in the NNPDF3.1sx study (see
the discussion in Sect.~4.1 of Ref.~\cite{Ball:2017otu}), finding
compatible results.

The conclusion is that the treatment of charm in the vicinity of charm
threshold deserves a very careful analysis, as it depends on many
details, which is however beyond the scope of this paper.  What is
instead relevant for us and very important to notice is that when
$\ln(1/x)$ resummation is included the quality of the description of
the data is largely independent of the possible differences in the
construction of the charm cross section, as noticed also in
Ref.~\cite{Ball:2017otu}: this is another achievement of high-energy
resummation.

\section{The role of low-$x$ and low-$Q^2$ data}
\label{sec:low}

\begin{figure}[t]
  \begin{centering}
    \includegraphics[width=0.49\textwidth]{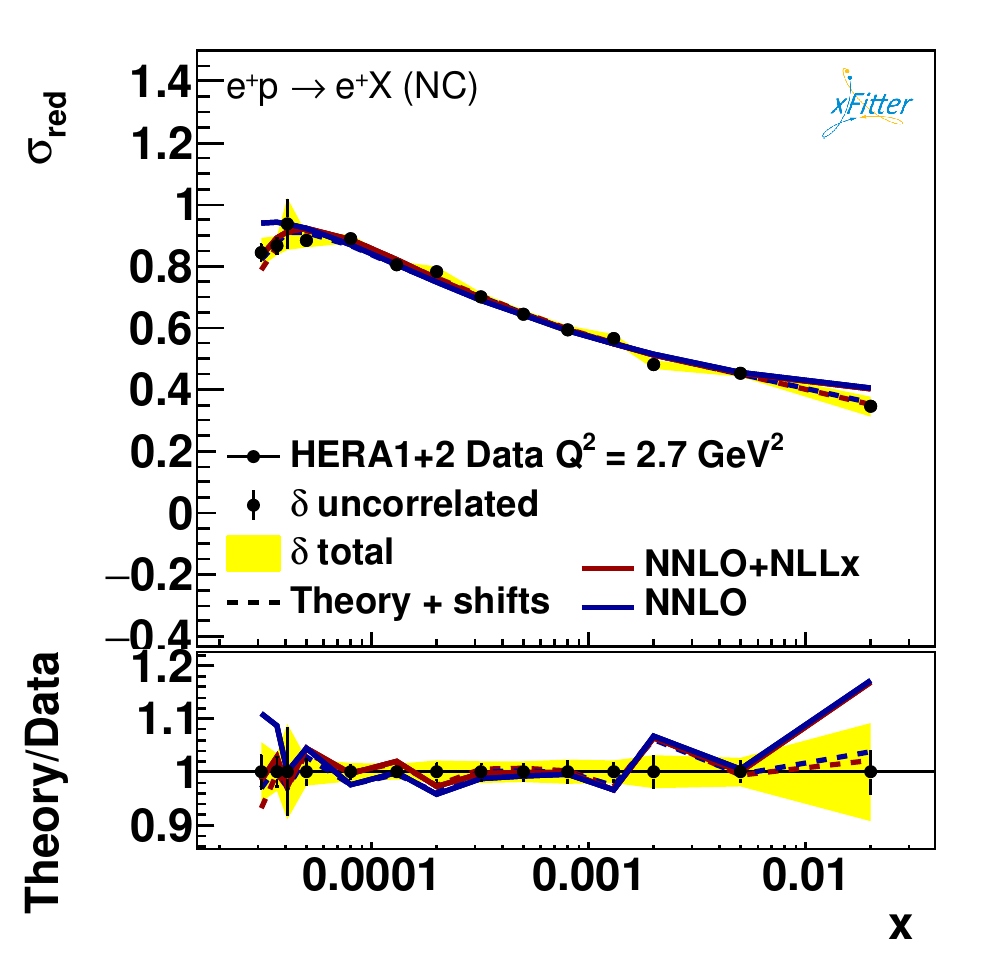}
    \caption{The HERA NC $E_p= 920$~GeV data at $Q^2 = 2.7$~GeV$^2$
      compared to the fits with and without $\ln(1/x)$ resummation
      including this bin.\label{fig:2pt5} }
  \end{centering}
\end{figure}

So far, we have maintained the restriction of the HERAPDF2.0 analysis,
keeping data with $Q^2\geq Q^2_{\rm min} = 3.5$~GeV$^2$. Since the
low-$Q^2$ data seem to be better described in the presence of
resummation, we can extend the fit down to lower values of $Q^2$ to
include the $Q^2=2.7$~GeV$^2$ bin of the $E_p=920$~GeV data
set,\footnote{Due to the limitation of {\tt HELL} at low scales, we
  cannot push the $Q^2_{\rm min}$ cut further down.} as was also done
in Ref.~\cite{Ball:2017otu}.  A visual inspection of
Fig.~\ref{fig:2pt5} shows that in the low-$x$ region the fit with
$\ln(1/x)$ resummation is able to describe these data points better
than the fixed-order fit. However, some discrepancies remain at large
$x$ that are not accommodated by resummation. The PDFs derived from
the fits including this extra $Q^2$ bin are very similar to those
shown in Fig.~\ref{fig:finalpdf} and are used to assess the model
uncertainty deriving from change in $Q^2_{\rm min}$ (however, note
that the upward variation to $Q^2_{\rm min}=5$~GeV has a significantly
larger impact on the shape of PDFs).  We will quantify the goodness of
the fits including this low-$Q^2$ bin below.

\begin{figure*}[t]
  \begin{centering}
    \includegraphics[width=0.32\textwidth]{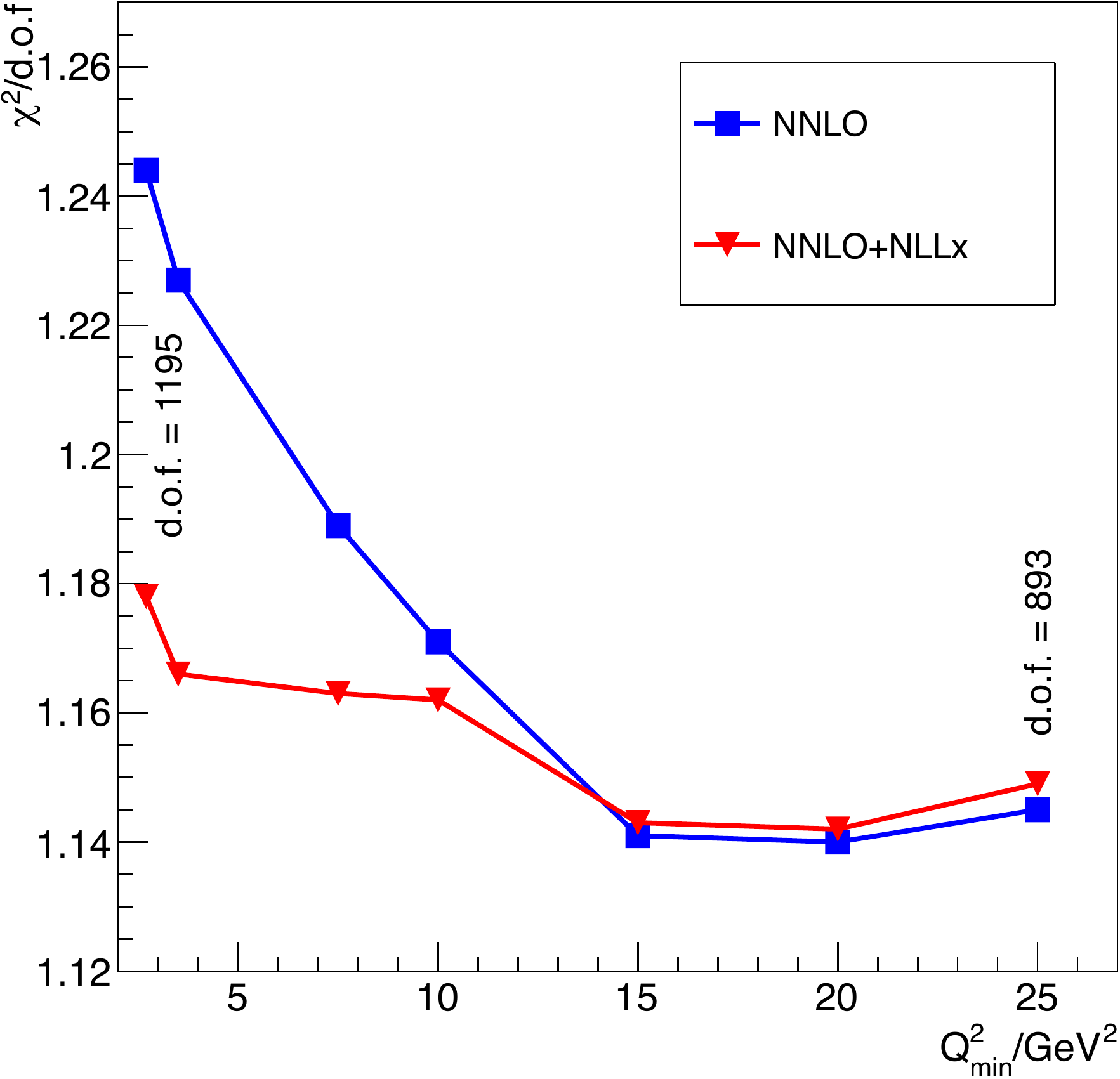}\hspace{5pt}
    \includegraphics[width=0.32\textwidth]{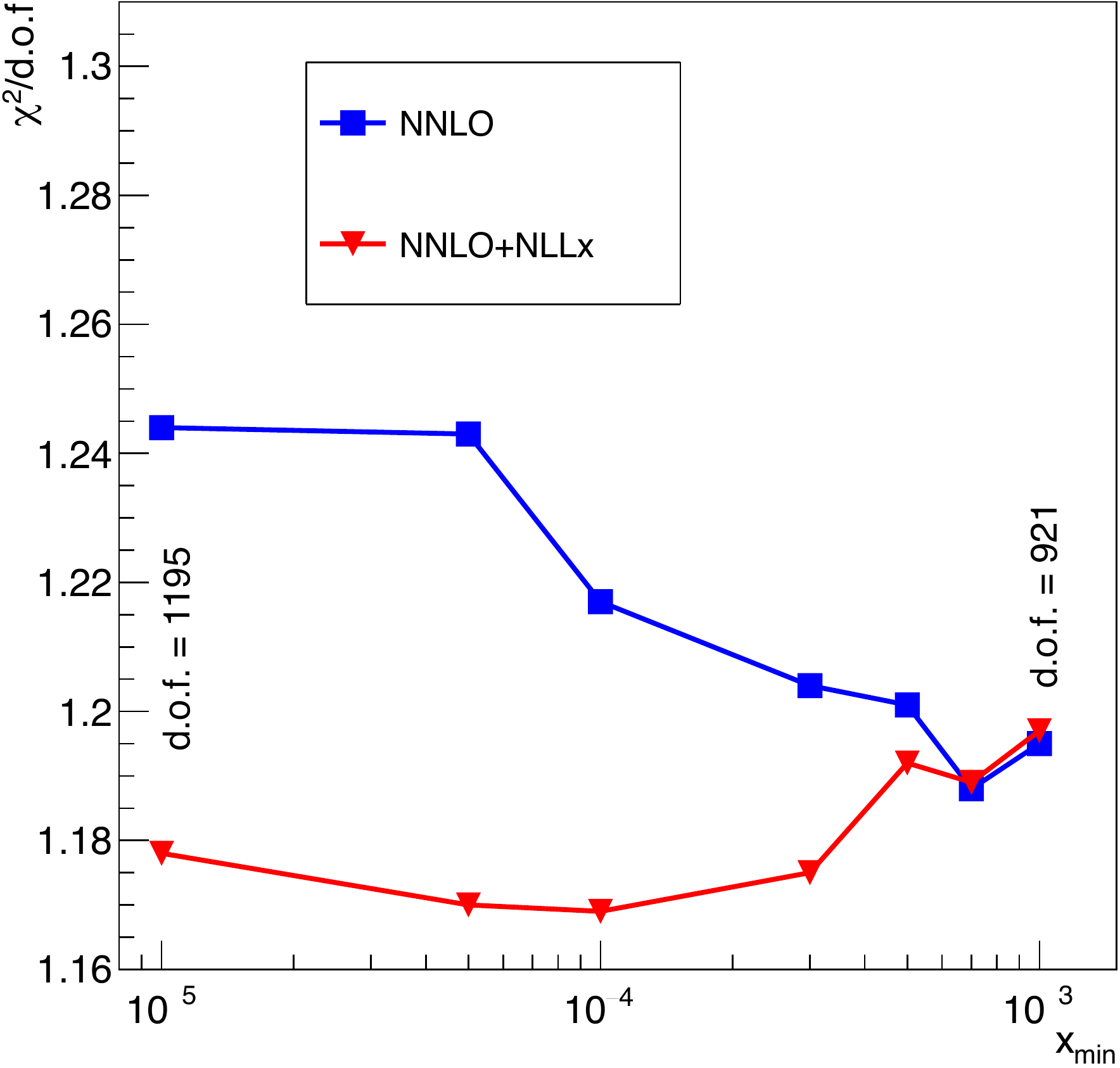}\hspace{5pt}
    \includegraphics[width=0.32\textwidth]{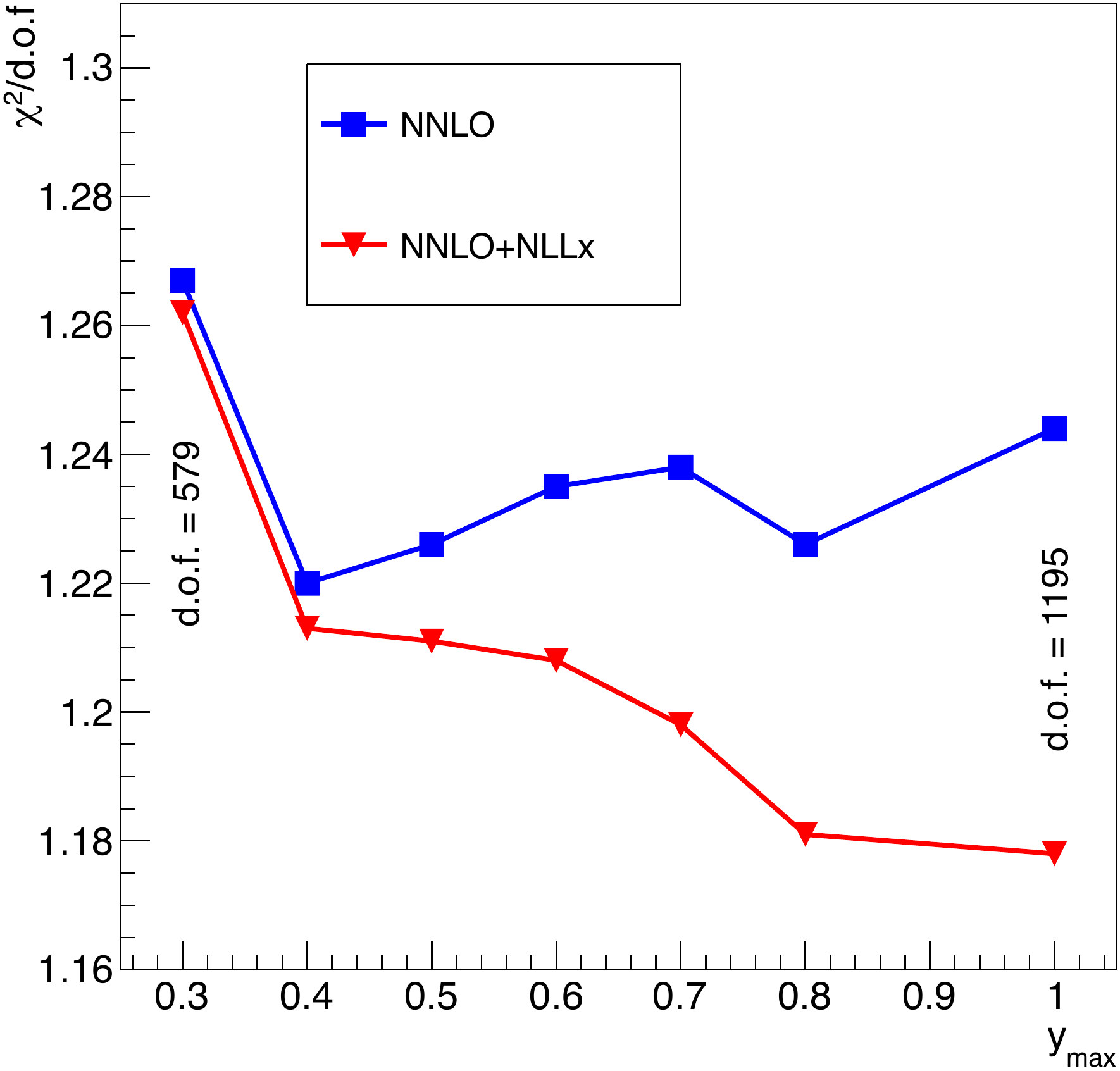}
    \caption{The $\chi^2$/d.o.f. as a function of $Q^2_{\rm min}$
      (left), $x_{\rm min}$ (center), and $y_{\rm max}$ (right). Each
      plot reports also the number of degrees of freedom at the
      extremities of each profile.}
    \label{fig:chi2profiles}
  \end{centering}
\end{figure*}

The results presented so far suggest that the improvement of the
description of the HERA data when including $\ln(1/x)$ resummation is
driven by the low-$x$ and low-$Q^2$ data. However, we can delineate
the kinematic region responsible for the improvement more
precisely. To do so, we have performed $\chi^2$ scans in
$Q^2_{\rm min}$ with no cut in $x$, and in $x_{\rm min}$ (where
$x_{\rm min}$ is the minumum value of Bjorken $x$ allowed in the fit)
fixing $Q^2_{\rm min} = 2.7$~GeV$^2$.\footnote {For these scans the
  beauty data are not included in the fits, since they have no impact
  as discussed in Sect.~\ref{sec:results}.} The results are shown in
Fig.~\ref{fig:chi2profiles} in the form of $\chi^2$/d.o.f.
profiles. From the $Q^2_{\rm min}$ scan (left plot) we observe that
$\ln(1/x)$ resummation provides a better description of the HERA data
from $Q^2_{\rm min} = 2.7$~GeV$^2$ up to
$Q^2_{\rm min}\simeq 15$~GeV$^2$, where resummed and fixed-order fits
converge towards the same $\chi^2$ values. The $x_{\rm min}$ scan
(central plot) shows that $\ln(1/x)$ resummation is significantly
better than fixed order up to $x_{\rm min}\simeq 5\cdot 10^{-4}$.
This allows us to conclude that $\ln(1/x)$-resummation effects improve
the description of the HERA data in the region
$x\lesssim 5\cdot 10^{-4}$ and $Q^2\lesssim 15$~GeV$^2$.

As mentioned above, a significant part of the improvement observed in
the low-$Q^2$ and low-$x$ region comes from an improved description of
$F_L$. Since $F_L$ contributes to the reduced cross section through a
factor $-y^2/Y_+$ (see Eq.~\eqref{eq:redxsec}), it is instructive to
do an additional $\chi^2$ scan in $y_{\rm max}$, excluding from the
fit data with $y > y_{\rm max}$. The $\chi^2$/d.o.f. as a function of
$y_{\rm max}$ is shown in the right plot of
Fig.~\ref{fig:chi2profiles}. Note that in this scan we set
$Q^2_{\rm min} = 2.7$~GeV$^2$ while no cut on $x_{\rm min}$ is
imposed. As expected, the $\chi^2$ profiles are very similar for small
values of $y_{\rm max}$ while they start diverging for
$y_{\rm max}\gtrsim 0.4$.

In the $\chi^2$ scans discussed above, full PDF fits were performed
for each different cut. In addition to this, we have performed studies
in which the $\chi^2$ are simply re-evaluated for the same value of
$x_{\rm min}$, $Q^2_{\rm max}$ and $y_{\rm max}$ using fixed PDFs,
specifically those for the NNLO and NNLO+NLL$x$ fits with
$Q_{\rm min}^2 = 2.7$~GeV$^2$ and no additional cuts in $x$ and $y$.
Similar trends as those found when refitting were observed. Finally,
we have performed a $\chi^2$ scan and a $\chi^2$ re-evaluation using
the variable $H = \ln (1/x) / \ln (Q^2/ \Lambda^2) $ with
$\Lambda = 88$~MeV defined in Ref.~\cite{Ball:2017otu}. We find that
the low-$H$ region is described similarly well by both NNLO and
NNLO+NLL$x$ fits while at high $H$ the $\ln(1/x)$ resummation improves
the data description. This is in agreement with the findings of
Ref.~\cite{Ball:2017otu}.

\begin{figure}[t]
  \begin{centering}
    \includegraphics[width=0.49\textwidth]{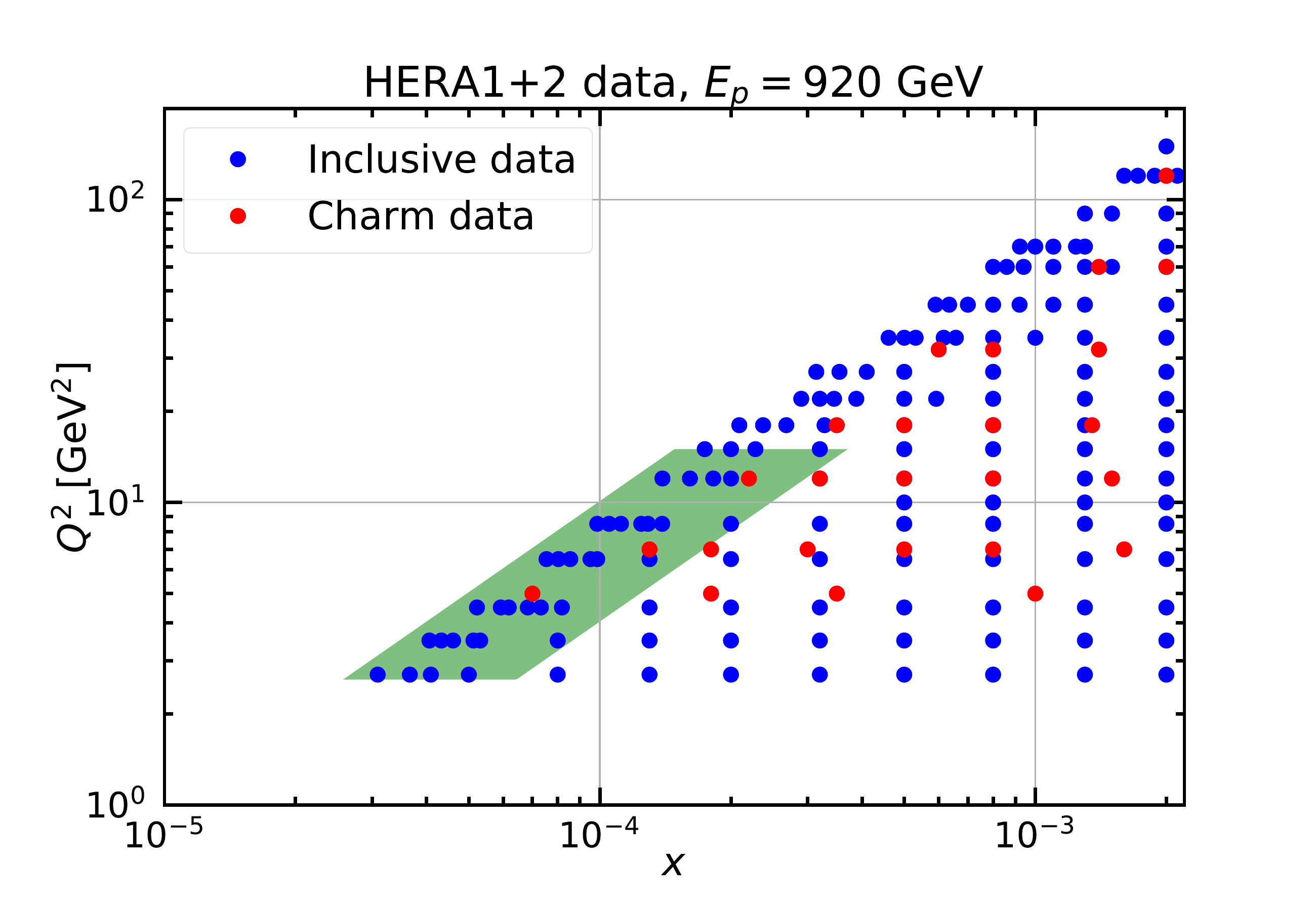}
    \caption{Scatter plot of the low-$x$ and low-$Q^2$ kinematic
      region covered by the HERA1+2 inclusive data and charm data at
      $E_p= 920$~GeV. The green shaded area indicates the region in
      which $\ln(1/x)$ resummation has a significant effect as
      compared to fixed order.\label{fig:scatterplot}}
  \end{centering}
\end{figure}
The $\chi^2$ scans as a function of $Q_{\rm min}$, $x_{\rm min}$ and
$y_{\rm max}$ allow us to delineate the region of the $(x,Q^2)$-plane
in which $\ln(1/x)$ resummation is important.\footnote{The actual
plane over which the constraint acts is the
$(x,Q^2/s)$-plane. However, for simplicity, in the following we will
only consider the $E_p=920$~GeV inclusive and the charm data sets that
were both taken at $\sqrt{s}=318$~GeV.}
Fig.~\ref{fig:scatterplot} displays a zoom of the low-$x$ and
low-$Q^2$ kinematic region covered by the HERA1+2 inclusive and charm
data at $E_p= 920$~GeV. The green shaded area indicates the region
such that $x<5\cdot 10^{-4}$, 2.7~GeV$^2 < Q^2 < 15$~GeV$^2$, and
$0.4 < y < 1$ (assuming $\sqrt{s} = 318$~GeV) determined by combining
the results of the scans discussed above.\footnote{In fact, given the
  ranges in $Q^2$ and $y$, the constraint on $x$ has no effect on the
  shaded area.} This provides an estimate of the region where
$\ln(1/x)$ resummation provides a significantly better description of
the HERA data as compared to fixed order. Since the $\chi^2$ scans in
Fig.~\ref{fig:scatterplot} have been obtained independently from one
another, one may wonder whether the estimate is fully reliable. In
order to check this, we have performed two additional fits, one with
and one without resummation, excluding only the data points for which
$Q^2<15$~GeV$^2$ and $y>0.4$. The total $\chi^2$'s of these fits
differ by around 15 units in favour of the resummed fit, mostly due to
the correlated and logarithmic terms, to be compared to the 73 units
of difference (see Tab.~\ref{tab:fitresults2}) when the shaded area is
included. This confirms that, in the context of DIS, the shaded area
in Fig.~\ref{fig:scatterplot} does provide a reliable estimate of the
kinematic region in which resummation works significantly better than
fixed order.

\section{Discussion and summary}

The recent implementation of the $\ln(1/x)$-resummation corrections to
the DGLAP splitting functions and the DIS coefficient functions in the
public code {\tt HELL}~\cite{Bonvini:2016wki,Bonvini:2017ogt} has made
possible the determination of PDFs including these effects. This
possibility has already been exploited in the recent global analysis
of Ref.~\cite{Ball:2017otu}. In this paper we focused on the study of
$\ln(1/x)$-resummation effects on the description of the HERA data in
the framework of an HERAPDF analysis. Specifically, we carried out a
PDF extraction from the HERA1+2 combined inclusive and charm
data~\cite{Abramowicz:2015mha,Abramowicz:1900rp} in the FONLL-C
variable-flavour-number scheme, accu\-rate to NNLO in QCD, including
and excluding resummation corrections up to NLL$x$ accuracy. This was
possible thanks to the {\tt xFitter} program~\cite{Alekhin:2014irh}
interfaced to the {\tt APFEL} code~\cite{Bertone:2013vaa}.

The inclusion of the $\ln(1/x)$-resummation effects makes the shape of
the gluon PDF at low $x$ and low scales steeply rising as opposed to
flattish/decreasing of the fixed-order fit (see
Fig.~\ref{fig:finalpdf}). The behaviour of the total singlet and gluon
PDFs towards low $x$ is much more similar when $\ln(1/x)$ resummation
is included and the ratio $\Sigma/g$ does not exceed unity in the
region of validity of the fit, $Q^2 > 2.56$~GeV$^2$. These features
make PDFs with $\ln(1/x)$ resummation much more suitable for use in MC
generators, such as Sherpa~\cite{Gleisberg:2008ta}, which require
positivity of the gluon distribution at all scales, than the standard
fixed-order NLO and NNLO PDFs, which have a suppressed gluon PDF at
low $x$ (however, for consistency, one should also include resummation
in the MC generators themselves).

The quality of the fit with $\ln(1/x)$ resummation is
si\-gni\-ficantly better than that of the corresponding fixed-order
analysis, indicating a better description of the HERA data. A
substantial part of the improvement in the description is driven by
$F_L$ which determines the behaviour of the DIS reduced cross section
at large values of $y$ (\textit{cfr.} Eq.~(\ref{eq:redxsec})). The
improvement is particularly significant at small values of $x$ and
$Q^2$ due to the relative size of the gluon PDF. In this region the
enhancement of $F_L$ caused by resummation helps reproduce the
turn-over of the data. The region of the $(x,Q^2)$-plane where
resummed predictions provide a better description of the HERA data was
delineated in Fig.~\ref{fig:scatterplot}.

In conclusion, $\ln(1/x)$ resummation provides a substantial
improvement in the description of the precise HERA1+2 combined data.
It represents an alternative to the addition of higher-twist
terms~\cite{Abt:2016vjh,Harland-Lang:2016yfn,Alekhin:2016uxn,Motyka:2017xgk}
and does not suffer from the pathological features of some of these
analyses~\cite{Abt:2016vjh}.  In addition, it overcomes a major
disadvantage of the fixed-order analyses, namely a decreasing gluon
PDF at low $x$ and $Q^2$.

\begin{acknowledgements}
  We would like to thank Juan Rojo for a cri\-ti\-cal reading of this
  paper, and Luca Rottoli for discussions on the description of charm
  data.  V.~B.\ and F.~G.\ are supported by the European Research
  Council Starting Grant ``PDF4BSM''.  Additional support was received
  by A.~G., A.~S.\ and P.~S.\ from the BMBF-JINR cooperation program
  and the Heisenberg-Landau program.  A.~L.\ is supported by the
  Polish Ministry under program Mobility Plus, no 1320/MOB/IV/2015/0.
  M.~B.\ is supported by the by the Marie Sk\l{}odowska Curie grant
  HiPPiE@LHC.
\end{acknowledgements}

\bibliographystyle{spphys}
\bibliography{lnxresum}

\begin{thebibliography}{10}
\providecommand{\url}[1]{{#1}}
\providecommand{\urlprefix}{URL }
\expandafter\ifx\csname urlstyle\endcsname\relax
  \providecommand{\doi}[1]{DOI \discretionary{}{}{}#1}\else
  \providecommand{\doi}{DOI \discretionary{}{}{}\begingroup
  \urlstyle{rm}\Url}\fi

\bibitem{Abramowicz:2015mha}
H.~Abramowicz, et~al., Eur. Phys. J. \textbf{C75}(12), 580 (2015).
\newblock \doi{10.1140/epjc/s10052-015-3710-4}

\bibitem{Gao:2017yyd}
J.~Gao, L.~Harland-Lang, J.~Rojo, arXiv:1709.04922  (2017)

\bibitem{CooperSarkar:1987ds}
A.M. Cooper-Sarkar, G.~Ingelman, K.R. Long, R.G. Roberts, D.H. Saxon, Z. Phys.
  \textbf{C39}, 281 (1988).
\newblock \doi{10.1007/BF01551005}

\bibitem{Abt:2016vjh}
I.~Abt, A.M. Cooper-Sarkar, B.~Foster, V.~Myronenko, K.~Wichmann, M.~Wing,
  Phys. Rev. \textbf{D94}(3), 034032 (2016).
\newblock \doi{10.1103/PhysRevD.94.034032}

\bibitem{Harland-Lang:2016yfn}
L.A. Harland-Lang, A.D. Martin, P.~Motylinski, R.S. Thorne, Eur. Phys. J.
  \textbf{C76}(4), 186 (2016).
\newblock \doi{10.1140/epjc/s10052-016-4020-1}

\bibitem{Alekhin:2016uxn}
S.~Alekhin, J.~Bluemlein, S.O. Moch, R.~Placakyte, PoS \textbf{DIS2016}, 016
  (2016)

\bibitem{Motyka:2017xgk}
L.~Motyka, M.~Sadzikowski, W.~Slominski, K.~Wichmann, arXiv:1707.05992  (2017)

\bibitem{Andreev:2013vha}
V.~Andreev, et~al., Eur. Phys. J. \textbf{C74}(4), 2814 (2014).
\newblock \doi{10.1140/epjc/s10052-014-2814-6}

\bibitem{Ball:2017otu}
R.D. Ball, V.~Bertone, M.~Bonvini, S.~Marzani, J.~Rojo, L.~Rottoli,
  arXiv:1710.05935  (2017)

\bibitem{Ball:1995vc}
R.D. Ball, S.~Forte, Phys.Lett. \textbf{B351}, 313 (1995).
\newblock \doi{10.1016/0370-2693(95)00395-2}

\bibitem{Ball:1997vf}
R.D. Ball, S.~Forte, Phys.Lett. \textbf{B405}, 317 (1997).
\newblock \doi{10.1016/S0370-2693(97)00625-4}

\bibitem{Altarelli:1999vw}
G.~Altarelli, R.D. Ball, S.~Forte, Nucl. Phys. \textbf{B575}, 313 (2000).
\newblock \doi{10.1016/S0550-3213(00)00032-8}

\bibitem{Altarelli:2000mh}
G.~Altarelli, R.D. Ball, S.~Forte, Nucl. Phys. \textbf{B599}, 383 (2001).
\newblock \doi{10.1016/S0550-3213(01)00023-2}

\bibitem{Altarelli:2001ji}
G.~Altarelli, R.D. Ball, S.~Forte, Nucl. Phys. \textbf{B621}, 359 (2002).
\newblock \doi{10.1016/S0550-3213(01)00563-6}

\bibitem{Altarelli:2003hk}
G.~Altarelli, R.D. Ball, S.~Forte, Nucl. Phys. \textbf{B674}, 459 (2003).
\newblock \doi{10.1016/j.nuclphysb.2003.09.040}

\bibitem{Altarelli:2005ni}
G.~Altarelli, R.D. Ball, S.~Forte, Nucl. Phys. \textbf{B742}, 1 (2006).
\newblock \doi{10.1016/j.nuclphysb.2006.01.046}

\bibitem{Ball:2007ra}
R.D. Ball, Nucl. Phys. \textbf{B796}, 137 (2008).
\newblock \doi{10.1016/j.nuclphysb.2007.12.014}

\bibitem{Altarelli:2008xp}
G.~Altarelli, R.D. Ball, S.~Forte, PoS \textbf{RADCOR2007}, 028 (2007)

\bibitem{Altarelli:2008aj}
G.~Altarelli, R.D. Ball, S.~Forte, Nucl. Phys. \textbf{B799}, 199 (2008).
\newblock \doi{10.1016/j.nuclphysb.2008.03.003}

\bibitem{Salam:1998tj}
G.~Salam, JHEP \textbf{9807}, 019 (1998).
\newblock \doi{10.1088/1126-6708/1998/07/019}

\bibitem{Ciafaloni:1999yw}
M.~Ciafaloni, D.~Colferai, G.~Salam, Phys.Rev. \textbf{D60}, 114036 (1999).
\newblock \doi{10.1103/PhysRevD.60.114036}

\bibitem{Ciafaloni:2003rd}
M.~Ciafaloni, D.~Colferai, G.~Salam, A.~Stasto, Phys.Rev. \textbf{D68}, 114003
  (2003).
\newblock \doi{10.1103/PhysRevD.68.114003}

\bibitem{Ciafaloni:2007gf}
M.~Ciafaloni, D.~Colferai, G.~Salam, A.~Stasto, JHEP \textbf{0708}, 046 (2007).
\newblock \doi{10.1088/1126-6708/2007/08/046}

\bibitem{Thorne:1999sg}
R.S. Thorne, Phys. Lett. \textbf{B474}, 372 (2000).
\newblock \doi{10.1016/S0370-2693(00)00019-8}

\bibitem{Thorne:1999rb}
R.S. Thorne, Phys. Rev. \textbf{D60}, 054031 (1999).
\newblock \doi{10.1103/PhysRevD.60.054031}

\bibitem{Thorne:2001nr}
R.S. Thorne, Phys. Rev. \textbf{D64}, 074005 (2001).
\newblock \doi{10.1103/PhysRevD.64.074005}

\bibitem{White:2006yh}
C.D. White, R.S. Thorne, Phys. Rev. \textbf{D75}, 034005 (2007).
\newblock \doi{10.1103/PhysRevD.75.034005}

\bibitem{Luszczak:2016bxd}
A.~Luszczak, H.~Kowalski, Phys. Rev. \textbf{D95}(1), 014030 (2017).
\newblock \doi{10.1103/PhysRevD.95.014030}

\bibitem{Bonvini:2016wki}
M.~Bonvini, S.~Marzani, T.~Peraro, Eur. Phys. J. \textbf{C76}(11), 597 (2016).
\newblock \doi{10.1140/epjc/s10052-016-4445-6}

\bibitem{Bonvini:2017ogt}
M.~Bonvini, S.~Marzani, C.~Muselli, JHEP \textbf{12}, 117 (2017).
\newblock \doi{10.1007/JHEP12(2017)117}

\bibitem{Alekhin:2014irh}
S.~Alekhin, et~al., Eur. Phys. J. \textbf{C75}(7), 304 (2015).
\newblock \doi{10.1140/epjc/s10052-015-3480-z}

\bibitem{h1zeus:2009wt}
F.~Aaron, et~al., JHEP \textbf{1001}, 109 (2010).
\newblock \doi{10.1007/JHEP01(2010)109}

\bibitem{Aaron:2009kv}
F.D. Aaron, et~al., Eur. Phys. J. \textbf{C64}, 561 (2009).
\newblock \doi{10.1140/epjc/s10052-009-1169-x}

\bibitem{Abramowicz:1900rp}
H.~Abramowicz, et~al., Eur. Phys. J. \textbf{C73}(2), 2311 (2013).
\newblock \doi{10.1140/epjc/s10052-013-2311-3}

\bibitem{Aaron:2009af}
F.D. Aaron, et~al., Eur. Phys. J. \textbf{C65}, 89 (2010).
\newblock \doi{10.1140/epjc/s10052-009-1190-0}

\bibitem{Abramowicz:2014zub}
H.~Abramowicz, et~al., JHEP \textbf{09}, 127 (2014).
\newblock \doi{10.1007/JHEP09(2014)127}

\bibitem{Bertone:2013vaa}
V.~Bertone, S.~Carrazza, J.~Rojo, Comput. Phys. Commun. \textbf{185}, 1647
  (2014).
\newblock \doi{10.1016/j.cpc.2014.03.007}

\bibitem{Forte:2010ta}
S.~Forte, E.~Laenen, P.~Nason, J.~Rojo, Nucl. Phys. \textbf{B834}, 116 (2010).
\newblock \doi{10.1016/j.nuclphysb.2010.03.014}

\bibitem{Bertone:2017ehk}
V.~Bertone, et~al., Eur. Phys. J. \textbf{C77}(12), 837 (2017).
\newblock \doi{10.1140/epjc/s10052-017-5407-3}

\bibitem{Botje:2010ay}
M.~Botje, Comput. Phys. Commun. \textbf{182}, 490 (2011).
\newblock \doi{10.1016/j.cpc.2010.10.020}

\bibitem{Thorne:1997ga}
R.S. Thorne, R.G. Roberts, Phys. Rev. \textbf{D57}, 6871 (1998).
\newblock \doi{10.1103/PhysRevD.57.6871}

\bibitem{Thorne:2006qt}
R.S. Thorne, Phys. Rev. \textbf{D73}, 054019 (2006).
\newblock \doi{10.1103/PhysRevD.73.054019}

\bibitem{Belov:2014xwo}
P.~Belov, et~al., Eur. Phys. J. \textbf{C74}(10), 3039 (2014).
\newblock \doi{10.1140/epjc/s10052-014-3039-4}

\bibitem{Aaron:2012qi}
F.D. Aaron, et~al., JHEP \textbf{09}, 061 (2012).
\newblock \doi{10.1007/JHEP09(2012)061}

\bibitem{Ball:2015tna}
R.D. Ball, V.~Bertone, M.~Bonvini, S.~Forte, P.~Groth~Merrild, J.~Rojo,
  L.~Rottoli, Phys. Lett. \textbf{B754}, 49 (2016).
\newblock \doi{10.1016/j.physletb.2015.12.077}

\bibitem{Ball:2015dpa}
R.D. Ball, M.~Bonvini, L.~Rottoli, JHEP \textbf{11}, 122 (2015).
\newblock \doi{10.1007/JHEP11(2015)122}

\bibitem{Ball:2017nwa}
R.D. Ball, et~al., Eur. Phys. J. \textbf{C77}(10), 663 (2017).
\newblock \doi{10.1140/epjc/s10052-017-5199-5}

\bibitem{Gleisberg:2008ta}
T.~Gleisberg, S.~Hoeche, F.~Krauss, M.~Schonherr, S.~Schumann, F.~Siegert,
  J.~Winter, JHEP \textbf{02}, 007 (2009).
\newblock \doi{10.1088/1126-6708/2009/02/007}

\end{thebibliography}

\end{document}